\documentclass[a4paper,11pt]{article}
\usepackage{graphicx}
\usepackage{fullpage}
\pdfoutput=1 

\usepackage{jheppub} 

\usepackage{bbm}
\usepackage{amsfonts}
\usepackage{slashed}
\usepackage{subfigure}
\usepackage{array}
\usepackage{verbatim}

\usepackage{booktabs}
\usepackage{color, soul}
\usepackage{mathrsfs}
\usepackage{epsfig}
\usepackage{graphicx}
\usepackage{dcolumn}
\usepackage{bm}
\usepackage{amsmath}
\usepackage{amssymb}

\usepackage{multirow}
\usepackage{dcolumn}

\title{\sffamily Interpreting the galactic center gamma-ray excess in the NMSSM}

\author[a,b]{Junjie Cao,}
\author[a]{Liangliang Shang,}
\author[c]{Peiwen Wu,}
\author[c]{Jin Min Yang,}
\author[c]{Yang Zhang}

\affiliation[a]{Department of Physics,
                Henan Normal University, Xinxiang 453007, China}
\affiliation[b]{Department of Applied Physics, Xi'an Jiaotong University, Xi'an 710049, China}
\affiliation[c]{State Key Laboratory of Theoretical Physics, Institute of Theoretical Physics,
                Academia Sinica, Beijing 100190, China}

\emailAdd{junjiec@itp.ac.cn}
\emailAdd{shlwell1988@gmail.com}
\emailAdd{pwwu@itp.ac.cn}
\emailAdd{jmyang@itp.ac.cn}
\emailAdd{zhangyang@itp.ac.cn}

\abstract{
In the Next-to-Minimal Supersymmetric Standard Model (NMSSM), all singlet-dominated particles including one neutralino, one CP-odd Higgs
boson and one CP-even Higgs boson can be simultaneously lighter than about 100 GeV. Consequently, dark matter (DM) in the
NMSSM can annihilate into multiple final states to explain the galactic center gamma-ray excess (GCE).  In this work
we take into account the foreground and background uncertainties for the GCE and investigate these explanations.
We carry out a sophisticated scan over the NMSSM parameter space by considering various experimental constraints
such as the Higgs data, $B$-physics observables, DM relic density, LUX experiment and the dSphs constraints. Then for each surviving
parameter point we perform a fit to the GCE spectrum by using the correlation matrix that incorporates both the
statistical and systematic uncertainties of the measured excess. After examining the properties of the obtained GCE solutions, we conclude
that the GCE can be well explained by the pure annihilations $\tilde{\chi}_1^0 \tilde{\chi}_1^0 \to b \bar{b} $ and
$\tilde{\chi}_1^0 \tilde{\chi}_1^0 \to A_1 H_i $ with $A_1$ being the lighter singlet-dominated CP-odd Higgs boson and
$H_i$ denoting the singlet-dominated CP-even Higgs boson or SM-like Higgs boson, and it can also be explained by the mixed
annihilation $\tilde{\chi}_1^0 \tilde{\chi}_1^0 \to W^+ W^-, A_1 H_1$. Among these annihilation channels,
$\tilde{\chi}_1^0 \tilde{\chi}_1^0 \to A_1 H_i $ can provide the best interpretation with the corresponding $p$-value reaching
0.55. We also discuss to what extent the future DM direct detection experiments can explore the GCE solutions and
conclude that the XENON-1T experiment is very promising in
testing nearly all the solutions.}

\begin{document}
\maketitle \indent
\newpage

\section{\label{intro}Introduction}

The compelling evidences for the existence of Dark Matter (DM) from various cosmological and astrophysical observations have provided us a good portal in the search for new physics beyond the Standard Model (SM). One possible method to explore DM in the present Universe is the indirect detection, which looks for the particles produced when DM annihilates in the DM halo. These particles include photons, antiparticles and neutrinos, and among them  gamma rays have often been
defined as the golden channel for DM indirect detection since the signal can be traced back to the source. The Large Area Telescope (LAT) onboard the Fermi Gamma-ray Space Telescope, due to its unprecedented angular and energy resolutions, has produced the most detailed maps of the gamma ray sky for a wide range of energies. Intriguingly, as was reported by several independent groups \cite{Goodenough_0910.2998,Hooper_1010.2752,Hooper_1110.0006, Abazajian_1207.6047,Gordon_1306.5725,Abazajian_1402.4090,Hooper_1302.6589,Hooper_1402.6703,Calore_1409.0042} and also by Fermi Collaboration itself  \cite{Murgia-Fermi},  the Fermi-LAT data have revealed the presence of an extended excess of gamma rays
over the modeled foreground and background emissions towards the Galactic Center (GC). Although several astrophysical mechanisms, such as the thousands of unresolved
millisecond pulsars  \cite{Yuan:2014rca,GCE-pulsars-Cholis:2014lta,O'Leary:2015gfa} and the interactions between
comic rays (CR) and interstellar gases \cite{Carlson:2014cwa,Petrovic:2014uda,Fields:2014pia,Zhou:2014lva,Gordon:2014gya},
have been proposed to interpret this Galactic Center Excess (GCE), they usually fail to generate the morphology and energy spectrum of
the GCE simultaneously\footnote{An exception may be the mechanisms recently proposed in  \cite{GCE-new-astro-2015-06}.}.
So in this work, we instead consider another possibility that the GCE is produced by the annihilation of DM.
Although this interpretation has been constrained by the measurements of CR such as the Fermi-LAT detection of the gamma-rays from dwarf spheroidal galaxies
(dSphs) \cite{Ackermann_1310.0828,Geringer-Sameth_1410.2242,Fermi-LAT-2014-Matthew-Wood,Fermi-LAT-2014-Anderson},
the non-observation of spectral features in the AMS-02 measurements of CR positron \cite{Bergstrom:2013jra,Ibarra:2013zia,Kong:2014haa,Yuan:2014pka},
and PAMELA observations of the CR anti-protons \cite{Kappl:2015bqa,Bringmann_1406.6027,Cirelli_1407.2173,Evoli:2011id,Cholis:2010xb,Donato:2008jk,Hooper_1410.1527},
it still remains a most attractive one not only because the excess emission shows spectral and morphological properties consistent with a telltale sign from DM annihilation,
but also because in such an interpretation, the annihilation cross section required to explain the GCE is of the right size to
account for the DM density from thermal freeze-out.

So far there have been a large number of attempts to explain the GCE by DM annihilation in various new physics models \cite{Kyae:2013qna,Modak:2013jya,Hardy:2014dea,Lacroix:2014eea, Alves:2014yha, Berlin:2014tja, Agrawal:2014una, Izaguirre:2014vva, Cerdeno:2014cda, Ipek:2014gua,
Ko:2014gha, Boehm:2014bia, Abdullah:2014lla, Ghosh:2014pwa, Martin:2014sxa, Berlin:2014pya, Basak:2014sza,
Cline:2014dwa, Han:2014nba, Wang:2014elb, Chang:2014lxa, Arina:2014yna, Cheung:2014lqa, Huang:2014cla,
Balazs:2014jla, Ko:2014loa, Baek:2014kna, Okada:2014usa, Bell:2014xta, Banik:2014eda, Borah:2014ska,
Cahill-Rowley:2014ora, Yu:2014pra, Guo:2014gra, Cao:2014efa, Yu:2014mfa, Freytsis:2014sua,
Heikinheimo:2014xza, Agrawal:2014oha, Cheung:2014tha, Petrovic:2014xra, Arcadi:2014lta, Hooper:2014fda,
Yuan:2014yda, Calore:2014nla, Liu:2014cma, Biswas:2014hoa, Ghorbani:2014gka, Cerdeno:2015ega,
Biswas:2015sva, Alves:2015pea, Kaplinghat:2015gha, Berlin:2015sia, Chen:2015nea, Guo:2015lxa,
Buckley:2015doa, Modak:2015uda, Caron:2015wda, Gherghetta:2015ysa, Elor:2015tva, Kopp:2015bfa,
Bi:2015qva, Appelquist:2015yfa, Rajaraman:2015xka,Cline:2015qha,Ko:2015ioa,Kim:2015fpa,Banik:2015aya}.
In the early analyses of the annihilations, great efforts were focused on the channels $\tilde{\chi} \tilde{\chi} \to b \bar{b}$ with $m_{\tilde{\chi}} \sim 35 {\rm GeV}$ and
$\tilde{\chi} \tilde{\chi} \to \tau \bar{\tau}$ with $m_{\tilde{\chi}} \sim 10 {\rm GeV}$ since they can reproduce well the GCE spectrum obtained at that time. Recently
a critical reassessment of the DM interpretation was made by examining in a comprehensive way the foreground and background uncertainties \cite{Calore_1409.0042}.
It was found that taking the estimated uncertainty in the high-energy tail of the spectrum into account, a much larger number of DM
annihilations are able to fit well the $\gamma$-ray data than previously noted \cite{Agrawal:2014oha,Calore:2014nla}. Explicitly speaking,
as far as the annihilation $\tilde{\chi} \tilde{\chi} \to b \bar{b}$ is concerned, now the mass of DM is extended to a broader range
from $30 {\rm GeV}$ to $70 {\rm GeV}$ in explaining the GCE \cite{Agrawal:2014oha,Calore:2014nla}. Other annihilation channels
such as DM annihilation into light quark pairs and even gluon pair are also able to provide a good fit to the GCE \cite{Calore:2014nla}.
More strikingly, this new analysis opens up a very good solution usually neglected before, namely DM annihilation into a
pair of light non-standard Higgs bosons \cite{Martin:2014sxa,Berlin:2014pya}. This important progress motivates us to
renew the solutions to the GCE in supersymmetric theories, which usually predict the lightest neutralino $\tilde{\chi}_1^0$ as a natural DM candidate.

As the most economical realization of supersymmetry, the Minimal Supersymmetric Standard Model (MSSM) is unsatisfactory in explaining the GCE due
to the following four reasons \cite{Cahill-Rowley:2014ora,Gherghetta:2015ysa}.
First, the relic density of DM has required its mass to be larger than about $40 {\rm GeV}$ \cite{Calibbi:2014coa}. In this case, the annihilations
$\tilde{\chi}_1^0 \tilde{\chi}_1^0 \to \tau \bar{\tau}, q \bar{q}$
with $q$ denoting a light quark can not provide a good fit any more. Second, except for excessive fine-tuning cases the LHC experiments have pushed
the lower mass bounds for the CP-odd Higgs boson and the bottom squarks up to several hundred ${\rm GeV}$. As a result, the cross section of DM
annihilation into $b\bar{b}$ in present day is too small to significantly contribute to the GCE \cite{Cahill-Rowley:2014ora,Gherghetta:2015ysa}.
Third, due to the small velocity of DM in our galaxy, the  annihilation rate for DM into SM-like Higgs pairs is $p$-wave suppressed.
Consequently this channel is not large enough to generate the GCE. Finally, as for the annihilations $\tilde{\chi}_1^0 \tilde{\chi}_1^0
\to WW, Z Z$, their fits to the GCE spectrum indicate that regardless of their annihilation rates the corresponding $p$-values are always less
than $0.04$ \cite{Agrawal:2014oha,Calore:2014nla}. This means that the annihilations can not generate the proper spectrum shape for the GCE.
We note that for a given parameter point of the MSSM, DM usually annihilates into multiple final states. In this case, the situation can not
be improved greatly because, due to the particle spectrum of the MSSM allowed by the current experiments, either the total cross section falls short for the GCE, or
the dominant annihilation channel can not reproduce the GCE spectrum well \cite{Caron:2015wda}.

Given the problems of the MSSM, we consider to interpret the GCE in the Next-to-Minimal Supersymmetric Standard Model (NMSSM) with a $Z_3$ symmetry, which is
the simplest gauge singlet Higgs extension of the MSSM \cite{NMSSM_review_0910.1785}.  Distinguished from the MSSM,
the NMSSM predicts three singlet-dominated particles: one neutralino, one CP-even and one CP-odd Higgs bosons. These particles are rather special in
that all of them can be simultaneously lighter than about $100 \,{\rm GeV}$, and that the couplings for the interactions among themselves
are determined by the parameter $\kappa$, which alone is able to predict the right rates for some annihilation channels to
explain the GCE (see the following discussion). These features make the NMSSM with a singlet-dominated DM well suit to account for the GCE because, as we will show below,
some golden channels for the GCE need light particles to act as the DM, the mediator and/or the annihilation final state.

We note that the interpretations of the GCE in the NMSSM have been intensively discussed in~ \cite{Cheung:2014lqa,
Huang:2014cla,Cahill-Rowley:2014ora,Guo:2014gra,Cao:2014efa,Gherghetta:2015ysa}. However, in \cite{Cheung:2014lqa,
Huang:2014cla,Cahill-Rowley:2014ora,Guo:2014gra,Cao:2014efa} the authors did not consider the systematic uncertainties mentioned above.
As a result, the model parameter space they considered is much narrower than that of this work and the obtained conclusions were incomplete.
While for \cite{Gherghetta:2015ysa}, although the authors have taken the uncertainties into account, they considered
the parameter space characterized by a large $\lambda$ which is different from our discussion.

The aim of this work is to explore any possible solution to the GCE in the $Z_3$ NMSSM. For this end, we perform a sophisticated scan
over the model parameters by considering various experimental constraints such as the DM relic density, the Higgs data as well as the observation of dwarf galaxies.
We use the correlation matrix presented in  \cite{Calore_1409.0042} to include the systematic uncertainties on the GCE spectrum and only keep the parameter points that can
reproduce well the spectrum. In our study we mainly consider a singlino-like DM which is believed to interpret the GCE without excessive fine tuning.
As we will show below, the annihilation $\tilde{\chi}_1^0 \tilde{\chi}_1^0 \to H_i A_1$ with $H_i A_1$ denoting a scalar-pseudoscalar Higgs pair may
provide the best fit to GCE, and the canonical annihilation $\tilde{\chi}_1^0 \tilde{\chi}_1^0 \to b \bar{b}$ still remains a satisfactory solution except that
$m_{\tilde{\chi}_1^0}$ is now allowed to vary within a broader range. Moreover, it is interesting to see that the mixed annihilation
into $W^+ W^-$ and $H_i A_1$ final states is also able to generate a spectrum consistent with the GCE.
These conclusions are quite different from previous studies in the NMSSM.

This paper is organized as follows. In Section II, we introduce some of the characteristic features of NMSSM, the basic knowledge about the GCE and
our strategy for the parameter scan. In Section III, we discuss in detail the interpretations of the GCE  when $H_2$ is the SM-like Higgs boson,
and in Section IV, we carry out a similar study but for the case that $H_1$ acts as the SM-like Higgs boson.
We draw our conclusion in Section V and provide more information of the NMSSM couplings in the Appendix.

\section{Fitting the GCE in the NMSSM}

\subsection{Theoretical setup for the GCE in the NMSSM}

We start our analysis by recapitulating the basics of the NMSSM. As one of the most economical extensions of the MSSM,
the NMSSM introduces one gauge singlet Higgs superfield in its matter content, and since one purpose of the extension
is to solve the $\mu$-problem of the MSSM, a $Z_3$ symmetry is usually adopted in the construction of the
superpotential to avoid the appearance of parameters with mass dimension.  As a result,  the
superpotential of the NMSSM and the soft breaking  terms in Higgs sector are given by  \cite{NMSSM_review_0910.1785}
\begin{eqnarray}
  W^{\rm NMSSM} &=& W_F + \lambda\hat{H_u} \cdot \hat{H_d} \hat{S}
  +\frac{1}{3}\kappa \hat{S^3},\\
  V^{\rm NMSSM}_{\rm soft} &=& \tilde m_u^2|H_u|^2 + \tilde m_d^2|H_d|^2
  +\tilde m_S^2|S|^2 +( \lambda A_{\lambda} SH_u\cdot H_d
  +\frac{1}{3}\kappa A_{\kappa} S^3 + h.c.),
\end{eqnarray}
where $W_F$ is the superpotential of the MSSM without the $\mu$-term, $\hat{H_u}$, $\hat{H_d}$
and $\hat{S}$ are Higgs superfields with $H_u$, $H_d$ and $S$ acting as their scalar components respectively,
the dimensionless coefficients $\lambda$ and $\kappa$ parameterize the strengthes of the Higgs self couplings,
and $\tilde{m}_{u}$, $\tilde{m}_{d}$, $\tilde{m}_{S}$, $A_\lambda$ and $A_\kappa$ are soft-breaking
parameters. In practice, after the electroweak symmetry breaking the soft-breaking squared masses $\tilde{m}_{u}^2$, $\tilde{m}_d^2$
and $\tilde{m}_s^2$ are traded for $m_Z$, $\tan \beta \equiv v_u/v_d$ and $\mu \equiv \lambda v_s $ as theoretical inputs.

Due to the presence of the superfield $\hat{S}$, the NMSSM contains a singlino field which is the fermion component of $\hat{S}$,
and one more complex Higgs field $S$ compared to the MSSM.  As a result, the neutralino mass
eigenstates $\tilde{\chi}_i^0$ (with $i$ ranging from $1$ to $5$) are the mixtures of bino, wino, higgsinos and singlino,
and the CP-even (odd) Higgs mass eigenstates $H_i$ with $i=1,2,3$ ($A_i$ with $i=1,2$) are mixtures of the real (imaginary)
parts of $H_u$, $H_d$ and $S$. Throughout this paper, we assume the mass order $m_{\tilde{\chi}_1^0} < m_{\tilde{\chi}_2^0}
< \cdots < m_{\tilde{\chi}_5^0}$ for neutralinos, and $m_{H_1} < m_{H_2} < m_{H_3} $, $m_{A_1} < m_{A_2}$ for Higgs bosons.

There are three distinguished features in the NMSSM. One is that DM in the NMSSM may be either singlino-dominated or
bino-dominated. As expected, the properties of a singlino-dominated DM are quite different from those
of a bino-dominated DM, which makes the DM physics in the NMSSM much richer than that in the MSSM  \cite{Cao-Light-DM}.
Another feature is that, in the presence of a singlino-dominated DM with mass below $100 {\rm GeV}$, the singlet-dominated
CP-even and CP-odd Higgs bosons can be simultaneously lighter than about $100 {\rm GeV}$ \cite{Cao-Light-DM,Cao-Light-Higgs},
and the strengthes for the interactions among these particles are determined by the parameter $\kappa$ which may be as large as 0.1.
This feature, as we will show below,
makes the NMSSM with a singlino-dominated DM well suit to explain the GCE. In the appendix, we list the properties of these
particles used in our analysis.  The other feature is that either $H_1$ or $H_2$ in the NMSSM can act as
the SM-like Higgs boson \cite{Cao-NMSSM} and generally speaking, $H_2$ as the SM-like boson is
more attractive from phenomenological point of view and also from naturalness argument.

\begin{figure}[t]
\begin{center}
\includegraphics[width=12cm]{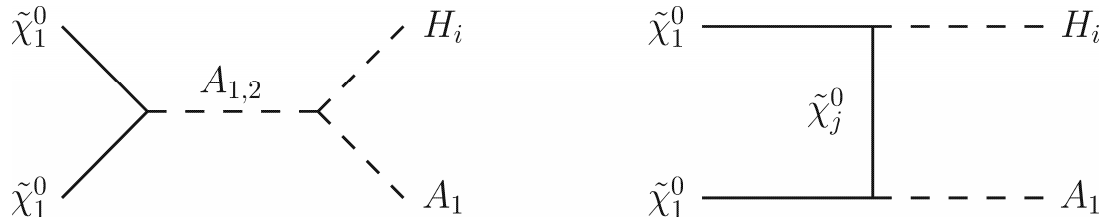}
\end{center}
\caption{Feynman diagrams contributing to the annihilation $\tilde{\chi}_1^0 \tilde{\chi}_1^0 \to H_i A_1$ with $\tilde{\chi}_j^0$ ($j$ from 1 to 5)  denoting any of the five neutralinos. A $u$-channel diagram in associated with the t-channel one is assumed. \label{fig1}}
\end{figure}

In the DM explanation of the GCE, the observed $\gamma$-ray originates mainly from the cascade decays of the annihilation final states.
In the NMSSM, the possible annihilation final states include $f\bar{f}$, $VV$, $H_i H_j$, $A_i A_j$ and $H_i A_j$ \cite{Griest-DM-annihilation},
where $f$ ($V$) denotes any of the fermions (vector bosons) in the SM, and $H_i$ ($A_j$) denotes a CP-even
(CP-odd) Higgs boson. In this work, we are particularly
interested in the annihilations $ \tilde{\chi}_1^0 \tilde{\chi}_1^0 \to b\bar{b}, W^+ W^-, H_i A_1$. These annihilations proceed
through the $s$-channel mediator of a Z boson or a Higgs boson with an appropriate CP quantum number, and also proceed through the $t/u$-channel
exchange of a sbottom, a chargino and a neutralino respectively. The complete expressions of the annihilation cross sections are rather
complicated, but in non-realistic limit, i.e. the velocity of DM approaching zero, some contributions become unimportant. In this case,
the velocity weighted annihilation cross section can be approximated by \cite{Griest-DM-annihilation}
\begin{eqnarray}
\langle \sigma_{b\bar{b}} v \rangle _{0} & \thickapprox &  \frac{3 \pi}{2} \sum_{i=1}^2 \frac{C_{A_i \tilde{\chi}_1^0 \tilde{\chi}_1^0}^2 C_{A_i b\bar{b}}^2 m_{\tilde{\chi}_1^0}^2}{(4m_{\tilde{\chi}_1^0}^2-m_{A_i}^2)^2+m_{A_i}^2\Gamma_{A_i}^2},   \label{bbar} \\
\langle \sigma_{WW} v \rangle_0 & \thickapprox & \frac{(\omega - 1 )^{3/2}}{32 \pi m_{\tilde{\chi}_1^0} m_W} \sum_{i=1}^2
\left ( \frac{f_{i,L}^2 + f_{i,R}^2}{1-\omega - k_i} \right )^2,  \label{WW} \\
 \langle \sigma_{H_i A_1} v \rangle_0  & \thickapprox &   \frac{1}{8\pi}\left(\frac{m_{h}}{m_{\tilde{\chi}_1^0}}\right)^{1/2}\left(1-\frac{m_{h}}{2m_{\tilde{\chi}_1^0}}\right)^{1/2} \sqrt{\delta}  \nonumber \\
 &&\hspace{-0.5cm}
 \times \left[\frac{C_{A_1 A_1 H_i} C_{A_1 \tilde{\chi}_1^0 \tilde{\chi}_1^0}}{m_{H_i}(4 m_{\tilde{\chi}_1^0}-m_{H_i})} +
 \frac{C_{A_2 A_1 H_i} C_{A_2 \tilde{\chi}_1^0 \tilde{\chi}_1^0}}{4 m^2_{\tilde{\chi}_1^0}-m^2_{A_2}} + 2\sum_{j=1}^{5}\frac{C_{A_1 \tilde{\chi}_1^0 \tilde{\chi}_j^0} C_{H_i \tilde{\chi}_1^0 \tilde{\chi}_j^0}}
 {m_{H_i}+|m_{\tilde{\chi}_j^0}|-m_{\tilde{\chi}_1^0}}\right]^2  \label{HIA1}
\end{eqnarray}
where $C_{XYZ}$ denotes the coupling of the interaction involving the particles $X$, $Y$ and $Z$,
$\Gamma_{A_i}$ is the width of the CP-odd state $A_i$, $\omega = m_{\tilde{\chi}_1^0}^2/m_W^2$,
$k_i = m_{\tilde{\chi}_i^\pm}^2/m_W^2$, $f_{i,L}$ ($f_{i,R}$) is the coupling
coefficient for $\tilde{\chi}_1^0 \tilde{\chi}_{i, L}^\pm W^\mp$ ($\tilde{\chi}_1^0 \tilde{\chi}_{i, R}^\pm W^\mp$) interaction, and
$\delta \equiv (2 m_{\tilde{\chi}_1^0} - (m_{H_i} + m_{A_1}))/2 m_{\tilde{\chi}_1^0}$. In getting Eq.(\ref{HIA1}),
we note that a good fit to the GCE requires that the $H_i A_1$ final state is produced
close to threshold, i.e. $\delta \simeq 0$, so we can expand  $\langle \sigma_{H_i A_1} v \rangle_0$
in terms of $\delta$. Then the first two terms on the right hand of  Eq.(\ref{HIA1}) come from the
left diagram of Fig.\ref{fig1}, and the last term comes from the right diagram of Fig.\ref{fig1}.

The flux per unit solid angle at some photon energy $E_{\gamma}$, which is observed by Fermi-LAT, is then given by
\begin{equation}
    \frac{{\rm d}\Phi_{\gamma}(E_{\gamma})}{{\rm d}E_{\gamma}{\rm d} \Omega} = \sum_{XY}
    \frac{\langle \sigma_{XY} v \rangle_0}{8\pi m_{\tilde{\chi}_1^0}^2}\frac{{\rm d}N_{XY}^{\gamma}}{{\rm d}E}
    \int{\rm ds} \: \rho_{\rm DM}^2(r(s \: , \: \theta))\;,  \label{flux}
\end{equation}
where $dN_{XY}^{\gamma}/dE$ is the photon spectrum generated by the annihilation $\tilde{\chi}_1^0 \tilde{\chi}_1^0 \to X Y$,
$\rho_{DM}$ is the DM profile and the integral over $\rho^2_{DM}$ is along the light-of-sight (LOS) at an angle $\theta$
towards GC. In the DM interpretation of the GCE,  a generalized Navarro,
Frenk \& White (NFW) DM profile is usually adopted, and its expression is given by~ \cite{Navarro:1995iw,Klypin:2001xu}
\begin{equation}
\rho(r)=\rho_\odot \left(\frac{r}{r_\odot}\right)^{-\gamma}\left(\frac{1+r_\odot/R_s}{1+r/R_s}\right)^{3-\gamma}
\label{genNFW}
\end{equation}
with slope parameter $\gamma=1.26$, scale radius $R_s=20$ kpc and the local DM density $\rho_\odot=0.4\, \mathrm{GeV}/\mathrm{cm}^3$
at the radial distance of the sun from the galactic centre $r_\odot$. Here the coordinate $r$ is centered on the galactic centre and
can be expressed as $r^2(s,\theta)=r_\odot^2+s^2-2 r_\odot s \cos\theta$ with $s$ and $\theta$ being the LOS distance
and the aperture angle between the axis connecting the earth with the galactic centre and the LOS respectively.

In our study, we use the package $\textsf{micrOMEGAs-3.6.9.2}$ \cite{micromegas} to calculate the DM relic density
and with the help of $\textsf{PYTHIA}$ \cite{Sjostrand:2014zea} to generate the flux in Eq.(\ref{flux}).
Note that in any explicit model, DM usually annihilates into multiple final states. In this case, the different fluxes are summed over.

\subsection{Parameter scan strategy for GCE solution}

We simplify our scan over the NMSSM parameter space by fixing the parameters that are not closely related to the DM studies.
The soft SUSY breaking parameters in the squark sector are all fixed to be 2 TeV except that
we vary those for the third generation to generate a CP-even Higgs near 125 GeV. We assume $A_t=A_b$ and $M_{U_3}=M_{D_3}$ to
reduce the number of free parameters. Similarly, all of the soft SUSY breaking parameters in the slepton sector are fixed to be 300 GeV
to explain the discrepancy of the measured value for muon anomalous magnetic moment from its SM prediction.
As for the gaugino sector we abandon the Grand Unified Theory relation and fix the wino mass and gluino mass
at $1 \,{\rm TeV}$ and $2 \,{\rm TeV}$ respectively. Consequently, the remained free parameters include
$\tan\beta, \mu, \lambda, \kappa, A_\lambda, A_\kappa$ in the Higgs sector, $M_{Q_3}$, $M_{U_3}$ and
$A_t$ for third generation quarks and the bino mass $M_1$, which are all defined at the scale of 2 TeV
in the scan. We use $\textsf{NMSSMTools-4.3.0}$ \cite{nmssmtools} to scan intensively the following NMSSM parameter region:
\begin{eqnarray}
&& 1 < \tan\beta < 40, ~0 < \lambda < 0.7, ~ 0 < |\kappa| < 0.7, ~|M_1| < 600 ~{\rm GeV},   \nonumber\\
&& 0 < A_\lambda < 5 ~{\rm TeV}, ~ |A_\kappa| < 2 ~{\rm TeV}, ~ |A_t| < 5 ~{\rm TeV}, \nonumber\\
&&  100~{\rm GeV}< \mu < 600 ~{\rm GeV},~ 200 ~{\rm GeV} < m_{Q_3}, m_{U_3} < 5 ~{\rm TeV}.
\end{eqnarray}

\begin{table}[th]
\begin{center}
\caption{Favored parameter region of the NMSSM to explain the GCE, which are classified by the dominant final state in DM annihilations.
These annihilations are called Solution I, II, III, IV and V respectively in the following discussion. All input parameters are defined at $2 {\rm TeV}$ and quantities with mass (annihilation cross section) dimension are in unit of GeV ($10^{-26} \, \rm{cm^3/s}$).}
\label{Table1}
\footnotesize
\vspace{0.3cm}
\begin{tabular}{|l|l|l|l|l|l|}
\hline
\multirow{2}{*}{}      & \multicolumn{3}{c|}{$H_2$ is SM-like}                     & \multicolumn{2}{c|}{$H_1$ is SM-like} \\ \cline{2-6}
                       & \multicolumn{1}{c|}{$b\bar{b}$} & \multicolumn{1}{c|}{$H_1A_1$} & \multicolumn{1}{c|}{$W^+W^-$} & \multicolumn{1}{c|}{$H_1A_1$} & \multicolumn{1}{c|}{$H_2A_1$} \\ \hline
$tan\beta$             & $(10,30)$         & $(8,40)$          & $(8,40)$          & $(15,20)$         & $(12,20)$         \\ \hline
$M_1$             & $(-600,-60)$      & $(-600,-80)$      & $(-600,-130)$     & $(-130,-90)$      & $(-200,-100)$     \\ \hline
$\lambda$              & $(0.2,0.7)$       & $(0.2,0.6)$       & $(0.2,0.4)$       & $(0.6,0.7)$       & $(0.4,0.7)$       \\ \hline
$\kappa$               & $(0.02,0.12)$     & $(0.07,0.15)$     & $(0.09,0.14)$     & $(0.10,0.14)$     & $(0.11,0.16)$     \\ \hline
$\mu$             & $(160,300)$       & $(110,210)$       & $(110,160)$       & $(220,270)$       & $(210,270)$       \\ \hline
$A_{\lambda}$     & $(2400,5000)$     & $(830,5000)$      & $(970,5000)$      & $(3900,5000)$     & $(2900,5000)$     \\ \hline
$A_{\kappa}$      & $(-210,-70)$      & $(-60,22)$        & $(-70,10)$        & $(-65,-16)$       & $(-66,5)$         \\ \hline
$A_{t,b}$     & $(-4300,3900)$    & $(-4600,4700)$    & $(-4700,3900)$    & $(-2200,2000)$    & $(-3400,4000)$    \\ \hline
$M_{Q_3}$         & $(300,5000)$      & $(350,5000)$      & $(500,5000)$      & $(1200,4600)$     & $(700,4800)$      \\ \hline
$M_{U_3,D_3}$     & $(250,5000)$      & $(270,5000)$      & $(400,5000)$      & $(250,5000)$      & $(1400,5000)$     \\ \hline
$m_{H_1}$         & $(15,102)$        & $(61,119)$        & $(83,110)$        & $(124,127.4)$     & $(124.5,127.4)$   \\ \hline
$m_{H_2}$         & $(122.8,127.8)$   & $(122.7,128)$     & $(123,128)$       & $(125.6,142)$     & $(125.7,146)$     \\ \hline
$\langle \sigma v \rangle _{0}$ & (0.17,1.9)      & (0.29,1.8)        & (0.44,1.6)        & (0.34,1.2)        & (0.38,1.5)        \\ \hline
$m_{\tilde{\chi}_1^0}$    & $(31,70)$         & $(62,114)$        & $(84,102)$        & $(71,87)$         & $(80,127)$        \\ \hline
$m_{\tilde{\chi}_2^0}$    & $(67,298)$        & $(83,233)$        & $(114,165)$       & $(86,128)$        & $(100,192)$       \\ \hline
$m_{\tilde{\chi}_1^{\pm}}$& $(166,297)$       & $(117,214)$       & $(117,158)$       & $(225,270)$       & $(218,266)$       \\ \hline
$m_{A_1}$         & $(58,133)$        & $(9,109)$         & $(10,105)$        & $(9,38)$          & $(16,95)$         \\ \hline
$m_{H^{\pm}}$     & $(3510,4666)$     & $(1477,3447)$     & $(2818,2968)$     & $(3883,4941)$     & $(2928,4740)$     \\ \hline
$\chi^2_{min}$ & $(23,35)$         & $(21,35)$         & $(24,35)$         & $(21,35)$         & $(21,35)$         \\ \hline
\end{tabular}
\end{center}
\end{table}

The process to retain the parameter points include the following steps:
\begin{itemize}
\item We require the DM to be singlino-dominated and satisfy $m_{\tilde{\chi}_1^0} \leq 150 ~{\rm GeV}$, and impose all the
experimental constraints encoded in $\textsf{NMSSMTools-4.3.0}$ \cite{nmssmtools} which include the relic
abundance at $3 \sigma$ level ($0.107\leq \Omega h^2 \leq 0.131$),
LUX exclusion bound at $90\%$ C.L., various B-physics measurements as well as the discrepancy of muon magnetic moment at $2\sigma$
level.  We also consider various electroweak precision data calculated in  \cite{Cao:2008rc}.
\item We consider the constraints on the Higgs sector with the package $\textsf{HiggsBounds-4.1.2}$ \cite{HiggsBounds}
which contains the data from LEP, Tevetron and LHC. For the SM-like Higgs boson, we further perform a fit to the data with the package
$\textsf{HiggsSignal}$ \cite{Bechtle:2013xfa} and keep the $2\sigma$ samples.
\item We use $\textsf{micrOMEGAs-3.6.9.2}$  \cite{micromegas} to calculate the DM annihilation cross section at present day,
and then impose the constraints from dSphs by the data in \cite{Fermi-LAT-2014-Anderson} for the $b\bar{b}$ annihilation channel and
with the method introduced in  \cite{Gherghetta:2015ysa} for the  $H_i A_1$ final states.
\item We also use $\textsf{micrOMEGAs-3.6.9.2}$ \cite{micromegas} to generate the $\gamma$-ray spectrum.
Considering the astrophysical uncertainties which may come from the errors in our setting on the local DM density $\rho_\odot$,
the scale radius $R_s$ and the inner slope parameter $\gamma$ in Eq.(7), for each parameter point we allow an uncertainty factor
$\mathcal{A}$ in the range of $(0.17,5.3)$ for the annihilation cross
section, or equivalently for the height of the gamma-ray spectrum in Eq.(\ref{flux})  \cite{Calore:2014nla}.
Then for the $\mathcal{A}-$tuned $\gamma$-ray spectrum, we perform a fit to the residual GCE spectrum obtained
in  \cite{Calore_1409.0042} by using the publicly available covariance matrix,  which include both the
statistical and systematic uncertainties of the measured flux.  The corresponding $\chi^2_{sp}$ function is
calculated by  \cite{Calore_1409.0042,Calore:2014nla}:
    \begin{equation}
	\chi^2_{sp}(\mathcal{A}) = \sum_{ij}\left(\frac{d\bar{N}}{dE_i} -\frac{dN}{dE_i}\right)
	\Sigma^{-1}_{ij}\left(\frac{d\bar{N}}{dE_j}-\frac{dN}{dE_j}\right),
	\label{eq:chi2}
    \end{equation}
where $\Sigma_{ij}$ is the covariance matrix, $dN/dE_i$ is the measured flux in the $i$-th energy bin, and $d\bar{N}/dE_i$
is the flux predicted by the NMSSM, which depends on the parameter point and also on the factor $\mathcal{A}$.

We define the GCE $\chi^2$ as the minimum value of $\chi^2_{sp}(\mathcal{A})$ among different choices of $\mathcal{A}$,
$\chi^2_{GCE} = min (\chi^2_{sp}(\mathcal{A}))$, and keep the parameter points that satisfy $\chi^2_{GCE} \leq 35.2$.
These points are assumed to have the capability to explain the GCE at $95\%$ confidence level
for 23 degree of freedom  \cite{Calore_1409.0042}.
\end{itemize}

The parameter ranges of the GCE solutions are listed in Table.\ref{Table1}, which are classified by the dominant
final state in DM annihilations (see the following discussion). For the first three types of the DM annihilations $H_2$ acts
as the SM-like Higgs boson, while for the last two types $H_1$ corresponds to the SM-like Higgs boson.
One distinguished feature that Table \ref{Table1} exhibits is that all the singlet dominated particles in the GCE solutions,
including DM, the singlet-dominated CP-even and CP-odd Higgs bosons, are lighter than about
$150 {\rm GeV}$. This feature, as we will emphasized below, makes the NMSSM well suit for explaining the GCE.

\begin{table}[th]
\caption{Detailed information of the benchmark points used in our discussion. Quantities with mass, annihilation and scattering cross section dimension are in unit of GeV, $\rm{cm^3/s}$ and pb respectively.}
\label{Table2}
\footnotesize
\begin{center}
\vspace{-0.5cm}
\begin{tabular}{|c|c|c|c|c|c|c|c|c|c|c|}
\hline
Point &~ $tan\beta$ ~&~~~ $\lambda$ ~~~&~~ $\kappa$ ~~&~~ $\mu$ ~~&~ $A_{\lambda}$ ~~&~~ $A_{\kappa}$ ~~&~ $A_{D_3,U_3}$ ~&~~ $M_1$ ~~&~~ $M_{Q_3}$ ~&~ $M_{U_3,D_3}$  \\ \hline
P1 & 16 & 0.36 & 0.04 & 241 & 3891 & -136 & 420   & -472 & 4127 & 4445 \\ \hline
P2 & 12 & 0.46 & 0.12 & 179 & 2036 & -6   & -2354 & -209 & 2197 & 3673 \\ \hline
P3 & 13 & 0.27 & 0.11 & 130 & 1899 & -5   & -524  & -170 & 4098 & 4384 \\ \hline
P4 & 18 & 0.69 & 0.12 & 243 & 4518 & -43  & -320  & -103 & 1436 & 4308 \\ \hline
P5 & 17 & 0.66 & 0.13 & 226 & 3923 & -17  & 1138  & -97  & 4540 & 1286 \\ \hline
P6 & 18 & 0.66 & 0.15 & 217 & 4048 & -24  & 2050  & -103 & 4170 & 1452 \\ \hline
P7 & 15 & 0.50 & 0.13 & 255 & 4085 & -35  & 2621  & -131 & 2935 & 4468 \\ \hline
\end{tabular}
\begin{tabular}{|c|c|c|c|c|c|c|c|c|c|c|}
\hline
Point  & $m_{H_1}$ & $m_{H_2}$ & $m_{H^{\pm}}$ & $m_{A_1}$ & $m_{\tilde{\chi}_1^0}$ & $m_{\tilde{\chi}_2^0}$ & $m_{\tilde{\chi}_1^{\pm}}$ & $Br_{(h2\to \tilde{\chi}_1^0\tilde{\chi}_1^0)}$ & $Br_{(h1\to A_1A_1)}$ & $Br_{(h2\to A_1A_1)}$ \\ \hline
P1 & 40  & 125 & 3960 & 99 & 50  & 256 & 248 & 0.54\%  & 0        & 0       \\ \hline
P2 & 99  & 125 & 2065 & 66 & 87  & 178 & 183 & 0       & 0        & 0       \\ \hline
P3 & 99  & 126 & 1823 & 48 & 92  & 126 & 134 & 0       & 88.51\%  & 7.06\%  \\ \hline
P4 & 126 & 133 & 4452 & 20 & 78  & 102 & 249 & 0       & 6.50\%   & 95.32\% \\ \hline
P5 & 125 & 126 & 3883 & 27 & 81  & 96  & 231 & 0       & 4.86\%   & 94.01\% \\ \hline
P6 & 126 & 129 & 4022 & 33 & 85  & 101 & 222 & 0       & 4.55\%   & 95.74\% \\ \hline
P7 & 125 & 145 & 4068 & 69 & 121 & 127 & 262 & 0       & 0        & 94.11\% \\ \hline
\end{tabular}
\begin{tabular}{|c|c|c|c|c|c|c|c|c|c|c|}
\hline
Point  & $\chi^2_{GCE}$ & p-value  & $\langle \sigma v \rangle _{0}$ & $\langle \sigma v \rangle |_{T_F}$ & $\sigma_p^{SI}$ & $\sigma_p^{SD}$ & $R_{b\bar{b}}$ & $R_{H_1A_1}$ & $R_{w^+w^-}$ & $R_{H_2A_1}$ ~\\ \hline
P1  & 23.3 & 0.44 & 6.1E-27   & 2.8E-26  & 2.3E-15 & 1.5E-04  & 90.4\% & 0.0\%  & 0.0\%  & 0.0\%  \\ \hline
P2 & 21.6  & 0.54 & 1.4E-26   & 2.6E-26  & 8.2E-10 & 1.3E-03  & 0.0\%  & 96.8\% & 3.1\%  & 0.0\%  \\ \hline
P3 & 24.8  & 0.36 & 1.3E-26   & 2.7E-26  & 1.9E-10 & 8.5E-04  & 0.1\%  & 46.1\% & 47.2\% & 0.0\%  \\ \hline
P4 & 22.4  & 0.50 & 8.6E-27   & 2.8E-26  & 5.5E-10 & 1.6E-03  & 0.1\%  & 94.3\% & 0.1\%  & 5.4\%  \\ \hline
P5 & 21.4  & 0.55 & 9.9E-27   & 3.1E-26  & 4.4E-10 & 1.7E-03  & 0.1\%  & 68.3\% & 0.2\%  & 31.4\% \\ \hline
P6 & 21.6  & 0.54 & 8.3E-27   & 3.2E-26  & 9.8E-10 & 2.0E-03  & 0.1\%  & 42.1\% & 2.1\%  & 55.5\% \\ \hline
P7 & 23.7  & 0.42 & 7.9E-27   & 3.3E-26  & 1.1E-09 & 5.5E-04  & 0.1\%  & 2.2\%  & 7.2\%  & 84.0\% \\ \hline
\end{tabular}
\end{center}
\end{table}

\section{GCE solutions with $H_2$ being the SM-like Higgs boson}

\begin{figure}[t]
\begin{center}
\includegraphics[width=10cm]{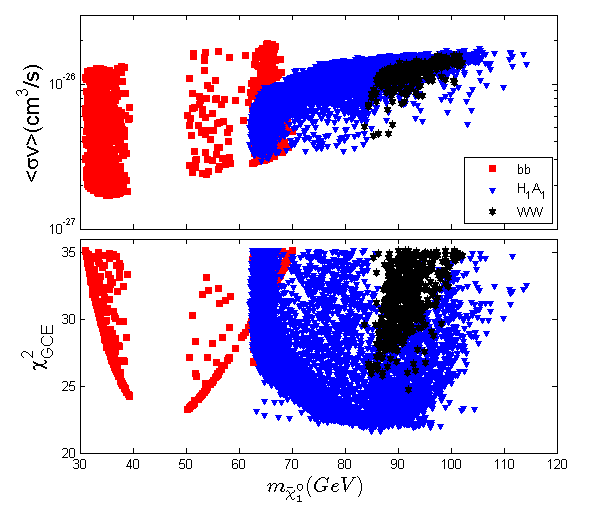}
\end{center}
\vspace{-.5cm}
\caption{The GCE solutions of the NMSSM for a singlino-dominated DM with $H_2$ acting as the SM-like Higgs boson, which are  projected
on the $\langle \sigma v \rangle_0-m_{\tilde{\chi}_1^0}$ plane (upper panel) and $\chi^2_{GCE}-m_{\tilde{\chi}_1^0}$ plane (lower panel).
Solutions marked by the red square, the blue triangle and the black asterisk correspond to the case that
DM annihilates in present day mainly by the channels $\tilde{\chi}_1^0 \tilde{\chi}_1^0 \to b \bar{b}, H_1 A_1, W^+ W^-$ respectively,
which are collectively called Solution I, II and III correspondingly. \label{fig2}}
\end{figure}

\begin{figure}[t]
\begin{center}
\includegraphics[width=14cm]{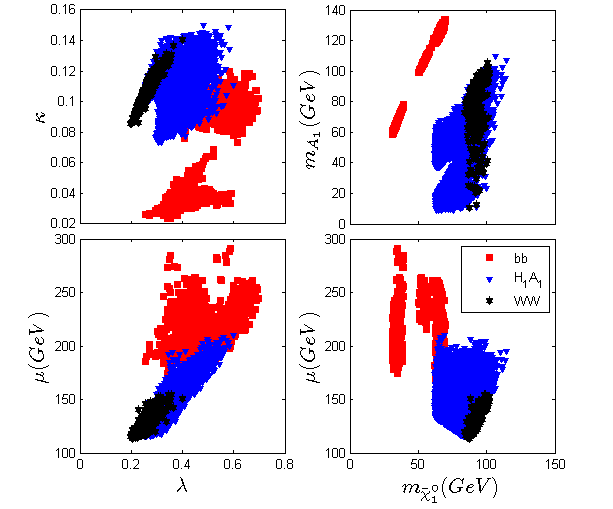}
\caption{Same as Fig.\ref{fig2}, but showing the correlations of different parameters. \label{fig3}}
\end{center}
\end{figure}

In this section, we exhibit the features of the GCE solutions for the case that DM is singlino-dominated and $H_2$ acts as the SM-like Higgs boson.
All the solutions considered in this work survive the constraints listed in last section and meanwhile can explain the GCE at $95\%$ C.L..

In Fig.\ref{fig1} we project the solutions on $\langle \sigma v \rangle_0-m_{\tilde{\chi}_1^0}$  plane (upper panel) and $\chi^2_{GCE}-m_{\tilde{\chi}_1^0}$ plane (lower panel). Solutions marked by red square, blue triangle and black asterisk correspond to the cases that DM annihilates with the largest branching ratio into $b\bar{b}$,
$H_1 A_1$ and $W^+ W^-$ final states respectively, which hereafter are collectively called Solution I, Solution II
and Solution III correspondingly.  Then the upper panel of  Figure.\ref{fig2} indicates that, for the ranges $30 {\rm GeV} \leq m_{\tilde{\chi}_1^0} \leq 40 {\rm GeV}$,
$50 {\rm GeV} \leq m_{\tilde{\chi}_1^0} \leq 62 {\rm GeV}$ and $63 {\rm GeV} \leq m_{\tilde{\chi}_1^0} \leq 70 {\rm GeV}$, Solution I is viable,
while for $63 {\rm GeV} \leq m_{\tilde{\chi}_1^0} \leq 115 {\rm GeV}$ and $83 {\rm GeV} \leq m_{\tilde{\chi}_1^0} \leq 100 {\rm GeV}$, Solution II and
Solution III can account for the GCE respectively. For any of the solutions, the $\langle \sigma v \rangle_0$ is larger than
$1.7 \times 10^{-27} {\rm cm^3/s}$, and its lower bound increases monotonically as $\tilde{\chi}_1^0$ becomes heavier. The reason for the
latter behavior is that, for a heavier DM, its number density is smaller. So to obtain the same photon flux for the GCE, a larger cross section is needed.

The lower panel of Fig.\ref{fig2} indicates that the best interpretation in Solution I comes from  $m_{\tilde{\chi}_1^0} \simeq 50 {\rm GeV}$
with $\chi^2_{GCE} \simeq  23$ and a $p$-value of 0.44.
This conclusion coincides with that of  \cite{Calore:2014nla}, which was obtained in a model independent way and for a pure $b\bar{b}$ annihilation
channel. For Solutions II and Solutions III, the best interpretations locate at $m_{\tilde{\chi}_1^0} \simeq 87 {\rm GeV}$ with $\chi^2_{GCE} \simeq  21.6$
and $m_{\tilde{\chi}_1^0} \simeq 92 {\rm GeV}$ with $\chi^2_{GCE} \simeq  24.7$ respectively,
and the corresponding p-values are 0.54 and 0.36. These two solutions, within our knowledge, were rarely
discussed in previous literatures about the NMSSM.  Moreover, we checked that, in the case of
$ m_{\tilde{\chi}_1^0} \simeq m_Z/2 $ ($ m_{\tilde{\chi}_1^0} \simeq 62 {\rm GeV} $), DM annihilated in early
universe mainly through a nearly on-shell $Z$ boson (SM-like Higgs boson). Since nowadays this dominant annihilation
channel is helicity ($p$-wave) suppressed, $\langle \sigma v \rangle_0$ can not reach the size required for the GCE.

In Table \ref{Table2}, we present detailed information of three benchmark points P1, P2 and P3 for Solution I, II and III respectively.
This table indicates that the sole annihilation channel $\tilde{\chi}_1^0 \tilde{\chi}_1^0 \to b \bar{b} $ or
$\tilde{\chi}_1^0 \tilde{\chi}_1^0 \to H_1 A_1 $ can be responsible for the GCE; while for the channel $\tilde{\chi}_1^0 \tilde{\chi}_1^0 \to W^+ W^- $,
it must mix sizeably with the channel  $\tilde{\chi}_1^0 \tilde{\chi}_1^0 \to H_1 A_1 $ to account for the GCE. We will return to this issue later.

In our calculation, we found that the condition on the GCE $\chi^2$ can reduce the number of the parameter points that survive the
constraints by more than $90\%$. This implies that the GCE has non-trivial requirements on the parameters of the
NMSSM, especially it suggests that some of the independent parameters may be correlated. Motivated by this thought,
we study the correlations among the parameters $\lambda$, $\kappa$, $\mu$, $m_{\tilde{\chi}_1^0}$ and
$m_{A_1}$ which are important parameters in the interpretation of the GCE and show the corresponding results in Fig.\ref{fig3}.
In the following, we concentrate separately on each kind of the solutions and investigate its features. Such a study is helpful to understand
the correlations in Fig.\ref{fig3} and also the properties of the benchmark points listed in Table.\ref{Table2}.

\subsection{Solution I - the $b \bar{b}$ annihilation channel}

Among the solutions to the GCE, Solution I is the most intensively studied one. After considering the systematic
uncertainties, one important improvement of Solution I over its previous version is that
DM mass is now allowed in the range from $30 {\rm GeV}$ to $70 {\rm GeV}$, which is much wider than before.

The key features of Solution I are as follows:
\begin{itemize}
\item The lighter CP-odd Higgs boson is correlated with DM by $m_{A_1} \simeq 2 m_{\tilde{\chi}_1^0}$. This correlation is
shown in the upper right panel of Fig.\ref{fig3} which means that the annihilation proceeds resonantly.

This feature can be understood as follows. In Solution I, the heavy CP-odd Higgs boson is doublet-dominated
with its mass usually at TeV scale. Then Eq.(\ref{bbar}) indicates
that the main contribution to the annihilation comes from the moderately light $A_1$, which is singlet-dominated.
With the formula presented in Eq.(\ref{couplings1}) and $v_s \equiv \mu/\lambda \gtrsim 450 {\rm GeV}$
shown in the lower left panel of Fig.\ref{fig3}, one can get
\begin{eqnarray}
C_{A_1 \tilde{\chi}_1^0 \tilde{\chi}_1^0}^2 C_{A_1 b\bar{b}}^2  ~ \simeq ~\lambda^2 \kappa^2 \left(\frac{m_b}{\mu}\right)^2
~ \lesssim ~ \left(\frac{5}{450}\right)^2 \kappa^2. \label{bbbar-approximate}
\end{eqnarray}
This inequation means that the couplings involved in the annihilation are highly suppressed so that the process must proceed
resonantly to ensure $\langle \sigma_{b\bar{b}} v \rangle _{0} \sim 10^{-26} cm^3/s$. Moreover, our results indicate that
the width of $A_1$ is very small, $\Gamma_{A_1} \lesssim 10^{-2} {\rm MeV}$. So as $m_{A_1}$ approaches  $2 m_{\tilde{\chi}_1^0}$,
the denominator in Eq.(\ref{bbar}) tends to vanish and a small  $\kappa$ in Eq.(\ref{bbbar-approximate})
is then suffice to predict the right rate of the annihilation for the GCE. This character is illustrated in the upper
left panel of Fig.\ref{fig3}. In fact, a small $\kappa$ is also favored to predict light $\tilde{\chi}_1^0$ and $A_1$,
which can be seen from Eq.(\ref{Basic-Approximation}) and Eq.(\ref{A1-mass}).

\item The parameter $\mu$ is upper bounded by about $300 {\rm GeV}$, which is shown in the
lower panels of Fig.\ref{fig3}.

This feature is actually required by the DM relic density \cite{Cao:2014efa}.  Generally speaking,  in order to predict the measured $\Omega h^2$,
the velocity weighted cross section $\langle \sigma v \rangle $ should be around the canonical value $3 \times 10^{-26} ~\rm{cm^3/s}$
at freezing out (see for example points in Table \ref{Table2}).  Since $2 m_{\chi_1^0}/m_{A_1} > 1$
in Solution I,  $\langle \sigma v \rangle $ for the annihilation
$\tilde{\chi}_1^0 \tilde{\chi}_1^0 \to A_1^\ast  \to b \bar{b} $ at present day
is usually larger than that at freezing out due to the thermal broadening \cite{Abnormal-annihilation}.
Since the dwarf galaxy measurements have required $\langle \sigma_{b\bar{b}} v \rangle_0 \lesssim 2 \times 10^{-26}\,\rm{cm^3/s}$
(see Fig.\ref{fig2}), new contributions such as those mediated by a $Z$ boson or a CP-even Higgs boson must intervene for the DM
annihilation in early Universe, and a moderately small $\mu$ can accelerate the annihilation  \cite{Cao:2014efa}.

\item Solution I suffers from severe fine tuning problem. Explicitly speaking, beside the correlation $m_{A_1} \simeq 2 m_{\tilde{\chi}_1^0}$,
there exits another strong correlation observed in our analysis, which is given by
\begin{eqnarray}
m_{\tilde{\chi}_1^0}/{\rm GeV}  \simeq  \left \{ \begin{array}{l}  51 - 475 \kappa, \quad for\ 30 {\rm GeV} \leq m_{\tilde{\chi}_1^0} \leq 40 {\rm GeV} \ or \ 0.024 \leq \kappa \leq 0.045, \nonumber  \\
37 + 325 \kappa, \quad for\ 50 {\rm GeV} \leq m_{\tilde{\chi}_1^0} \leq 62 {\rm GeV} \ or \ 0.038 \leq \kappa \leq 0.07, \nonumber  \\
49 + 175 \kappa, \quad for\ 63 {\rm GeV} \leq m_{\tilde{\chi}_1^0} \leq 70 {\rm GeV} \ or \ 0.08 \leq \kappa \leq 0.12. \nonumber
\end{array} \right.
\end{eqnarray}
These correlations make Solution I in the NMSSM quite unnatural to explain the GCE.

\item We checked that $Br(A_1 \to \gamma \gamma ) < 5 \times 10^{-4}$ so that the $\gamma$-ray spectral line generated by
$\tilde{\chi}_1^0 \tilde{\chi}_1^0 \to A_1^\ast  \to \gamma \gamma $ is suppressed.

\item Since $\tilde{\chi}_1^0 \lesssim 60 {\rm GeV}$ for most cases in Solution I, the SM-like Higgs boson $H_2$ may decay into
$\tilde{\chi}_1^0$ pair. We checked that $Br(H_2 \to \tilde{\chi}_1^0 \tilde{\chi}_1^0) \lesssim 18\%$, which is required by the Higgs
data at the LHC.
\end{itemize}

\subsection{Solution II - the $H_1 A_1$ annihilation channel}

Solution II is quite similar to the interpretations presented in~ \cite{Martin:2014sxa,Berlin:2014pya,
Cerdeno:2015ega,Cline:2015qha,Gherghetta:2015ysa,Ko:2015ioa},  which utilize the process $\tilde{\chi} \tilde{\chi} \to \phi_1 \phi_2 \to f_1 \bar{f}_1 f_2 \bar{f}_2$
($\phi_1$ and $\phi_2$ denote scalar or pseudoscalar particles, and $f_1$ and $f_2$ are SM fermions) for the GCE. These interpretations,
as were emphasized by the proposers, can easily escape the constraints from DM detection experiments and have been paid more and more attention recently.

The features of Solution II are as follows:
\begin{itemize}
\item The singlet-dominated particles satisfy $60 {\rm GeV} \lesssim  m_{\tilde{\chi}_1^0} \lesssim 115 {\rm GeV}$,
$10 {\rm GeV} \lesssim  m_{A_1} \lesssim 110 {\rm GeV}$, $60 {\rm GeV} \lesssim  m_{H_1} \lesssim 120 {\rm GeV}$ and $\delta < 0.2$,
and for most samples there exist following relations $m_{H_1} \geqslant m_{\tilde{\chi}_1^0} \geqslant m_{A_1}$.
Given $\kappa \sim 0.1 $ which is required to predict
the right size of the annihilation $\tilde{\chi}_1^0 \tilde{\chi}_1^0 \to H_1 A_1$ for the GCE (see below), the particle spectrum limits parameters such as
$\lambda$, $\mu$ and $A_\kappa$ in certain regions (see the expressions of the tree level masses in Appendix), which are given in Table \ref{Table1},
and also shown in Fig.\ref{fig3}.

Note that $\mu$ is below about $200 {\rm GeV}$. In this case, the higgsino-dominated neutralinos $\tilde{\chi}_i^0$ may decay dominantly into $ \tilde{\chi}_1^0 A_1 $
instead of into $ \tilde{\chi}_1^0 Z $ since the kinematics is forbidden. In this case, the LHC search for electroweakinos by trilepton $+ E_T^{miss}$ signal
is less efficient in ruling out the light higgsinos\footnote{In doing  \cite{Cao-Light-DM}, we once confronted with the situation quite similar to what
we are facing now. Our detailed simulation at that time indicated that the trilepton constraint on SUSY is very weak. Moreover, in comparison with the case discussed
in  \cite{Han-compressed}, we find that our case is more difficult to detect since the signal is smaller.}. Also note that the parameters
$\lambda$ and $\mu$ are related by $\mu/{\rm GeV} \thickapprox 60 + 260\  \lambda$ for $\lambda$ varying from 0.2 to 0.6 (see lower left panel of Fig.\ref{fig2}),
which means that $v_s \equiv \mu/\lambda > 360 {\rm GeV}$. This ensures that the expansions for the masses and
couplings in Appendix by the power of $\lambda v/\mu$ are good approximations.

\item The s-channel contributions to the annihilation rate $ \langle \sigma_{H_1 A_1} v \rangle_0 $ in Eq.(\ref{HIA1}) are usually much
smaller than those from the $t/u$ channel, and among the $t/u$ channel contributions, the one induced by the exchange of $\tilde{\chi}_1^0$
is far dominant. As for the contributions induced by the two higgsino-like neutralinos, each of them may be sizable,
but since they cancel each other, the net higgsino contribution is not important. These characters can be understood by
the following approximations (see Eq.(\ref{couplings1}))
\begin{eqnarray}
&& C_{A_1 \tilde{\chi}_1^0 \tilde{\chi}_1^0} C_{H_1 \tilde{\chi}_1^0 \tilde{\chi}_1^0} \simeq 2 i \kappa^2 ( 1 + 2 \frac{\lambda v}{\mu} )^2 , \nonumber \\
&& C_{A_1 \tilde{\chi}_1^0 \tilde{\chi}_i^0} C_{H_1 \tilde{\chi}_1^0 \tilde{\chi}_i^0} \simeq
\left \{ \begin{array}{c} - \frac{i}{4} \frac{\lambda^2 v^2}{\mu^2} \sin^2 \beta, \quad for\  Higgsino-like\ \tilde{\chi}_i^0\ and  \ m_{\tilde{\chi}_i^0} < 0, \nonumber  \\
\ \frac{i}{4} \frac{\lambda^2 v^2}{\mu^2} \sin^2 \beta, \quad for\  Higgsino-like\ \tilde{\chi}_i^0\ and  \ m_{\tilde{\chi}_i^0} > 0,  \nonumber
\end{array} \right.
\end{eqnarray}
and by the fact that $\kappa \sim 0.1$ is enough to predict the $\tilde{\chi}_1^0$ contributed $ \langle \sigma_{H_i A_1} v \rangle_0 $ at the order of $10^{-26} cm^3/s$
(see equation (3.20) in  \cite{Gherghetta:2015ysa}).

\item Since $m_{A_1} \lesssim 60 {\rm GeV}$ for most cases in Solution II (see upper right panel of Fig.\ref{fig2}), the SM-like Higgs boson $H_2$
may decay into $A_1 A_1$ with a sizeable fraction. Given that $A_1$ decays dominantly into $b \bar{b}$, this will result in $4b$ signal for
the SM-like Higgs boson.  We checked that $Br( H_2 \to A_1 A_1 ) \lesssim 24 \%$, where the upper bound comes from the constraints of the LHC Higgs data.

\item Since a good fit to the GCE requires that $H_1 A_1$ is produced close to threshold, the annihilation $\tilde{\chi}_1^0 \tilde{\chi}_1^0 \to H_1 A_1$
will produce spectral line or  box-shaped spectrum in  $\gamma$-ray \cite{Cerdeno:2015ega,Gherghetta:2015ysa}. We checked that
$Br(H_1 \to \gamma \gamma ) \leq 1 \times 10^{-3}$ for most samples and $Br(A_1 \to \gamma \gamma ) < 4 \times 10^{-4}$ for all samples.
So current results of the Fermi-LAT search for spectral lines \cite{Ackermann:2013uma} can not impose tight limit on Solution II (see  \cite{Cerdeno:2015ega} for
a detailed discussion).
\end{itemize}

\subsection{Solution III - the $W^+ W^-$ annihilation channel}

\begin{figure}[t]
\begin{center}
\includegraphics[width=12cm]{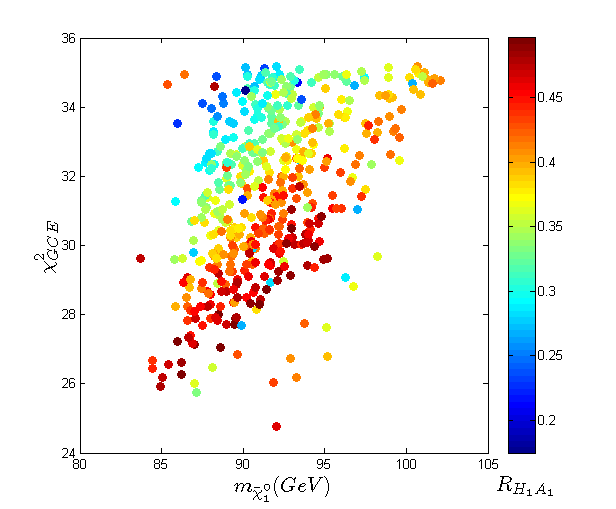}
\caption{The GCE $\chi^2$ as a function of DM mass for Solution III where DM annihilates mainly into $WW$ final state. Here $R_{H_1 A_1}$ denotes the branching ratio of the
annihilation into $H_1 A_1$ final state. \label{fig4}}
\end{center}
\end{figure}

In general, the pure annihilation $\tilde{\chi}_1^0 \tilde{\chi}_1^0 \to W^+ W^- $ is unable to explain the GCE quite well \cite{Agrawal:2014oha,Calore:2014nla},
but if it mixes sizably with other annihilation channels, the generated spectrum may be improved significantly to account for the GCE.
Solution III in the NMSSM belongs to this case.

Solution III has the following features:
\begin{itemize}
\item The $W$ pair must be produced close to threshold to account for the GCE, which means $85 {\rm GeV} \lesssim m_{\tilde{\chi}_1^0} \lesssim 100 {\rm GeV}$
(see right panels of Fig.\ref{fig3}).
\item From the expression of $\langle \sigma_{WW} v \rangle_0$ in Eq.(\ref{WW}), one can learn that, if the wino is decoupled, the annihilation rate is
determined by the higgsino-dominated chargino. In this case, we have
\begin{eqnarray}
f_{1, L} &\simeq & - \frac{g}{\sqrt{2}} N_{14} \simeq \frac{g}{\sqrt{2}} \frac{\lambda v}{\mu} \sin \beta \simeq \sqrt{2} g \sin \beta \frac{\kappa v}{m_{\tilde{\chi}_1^0}}, \nonumber \\
f_{1, R} &\simeq & - \frac{g}{\sqrt{2}} N_{13} \simeq - \sqrt{2} g \sin \beta \frac{\kappa v}{\mu}.
\end{eqnarray}
In getting these expressions, we note $v_s \equiv \mu/\lambda \gtrsim 400 {\rm GeV}$ (see lower left panel of Fig.\ref{fig2}), and expand $N_{13}$ and $N_{14}$
in terms of $\lambda v/\mu$ (see Appendix). We also use the approximation $m_{\tilde{\chi}_1^0} \simeq 2 \kappa \mu/\lambda$.
Then $\langle \sigma_{WW} v \rangle_0 \sim 10^{-26} cm^3/s$ and $m_{\tilde{\chi}_1^0} \sim 90 {\rm GeV}$ limit tightly the ranges of the parameters $\lambda$,
$\kappa$ and $\mu$, which are shown in Table \ref{Table1} and Fig.\ref{fig3}.

Note in Solution III, the parameter $\mu$, or equivalently the masses for the higgsino-dominated chargino and neutralinos, is less than about $150 {\rm GeV}$.
Since the splitting between $\mu$ and $m_{\tilde{\chi}_1^0}$ is less than about $50 {\rm GeV}$, such a low value of $\mu$ is still allowed by the
LHC search for SUSY (see footnote 2 in our discussion on Solution II).

\item The upper left panel of Fig.\ref{fig3} indicates that the parameters $\lambda$ and $\kappa$ are correlated by
\begin{eqnarray}
\kappa \simeq 0.03 + 0.3 \lambda, \quad for \ 0.2 \leq \lambda \leq 0.4.
\end{eqnarray}
As a result, we have $m_{\tilde{\chi}_1^0} \simeq 2 \mu/3$.

\item As we emphasized before, the annihilation $\tilde{\chi}_1^0  \tilde{\chi}_1^0 \to W^+ W^-$ must mix sizably with the
annihilation  $\tilde{\chi}_1^0  \tilde{\chi}_1^0 \to H_1 A_1$ to explain the GCE. This, in return, requires appropriate masses for $H_1$ and $A_1$
to improve the $\gamma$-ray spectrum generated by the $WW$ state.  In Fig.\ref{fig4},
we plot the GCE $\chi^2$ as a function of DM mass in Solution III with different colors denoting the branching ratio of the DM annihilation
into $H_1 A_1$. This figure indicates that, with the increase of the branching ratio, the GCE $\chi^2$ tends to decrease.
\end{itemize}

\section{GCE solutions with $H_1$ being the SM-like Higgs boson}

\begin{figure}[t]
\begin{center}
\includegraphics[width=14cm]{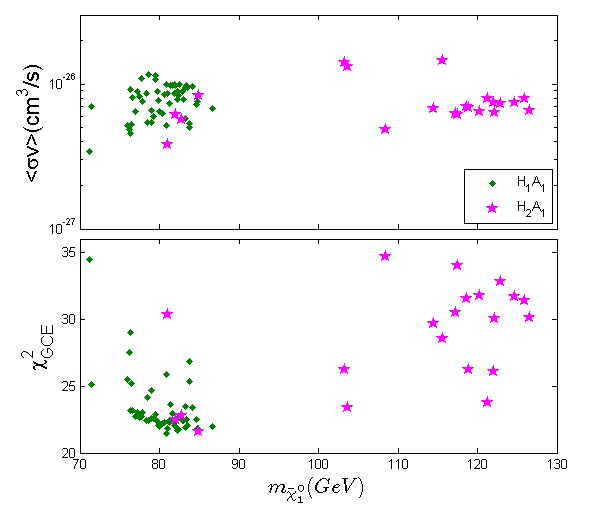}
\caption{Similar to Fig.\ref{fig2}, but showing the solutions for the case that $H_1$ acts as the SM-like Higgs boson. For these solutions,
DM may mainly annihilate into $H_1 A_1$ final state (called Solution IV in our discussion) or into $H_2 A_1$ final state (Solution V).
 \label{fig5}}
\end{center}
\end{figure}

\begin{figure}[t]
\begin{center}
\includegraphics[width=14cm]{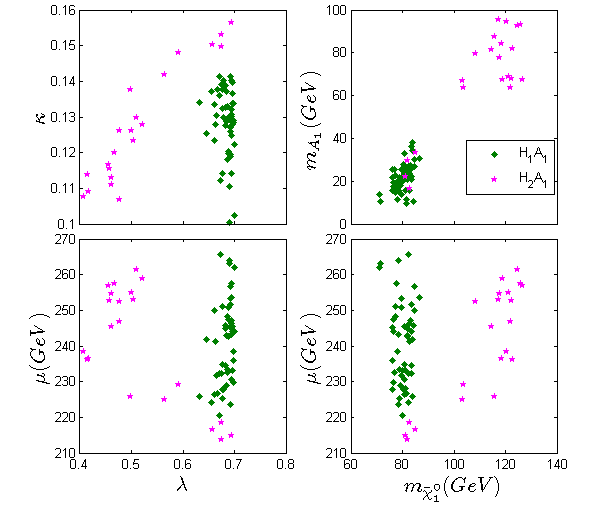}
\caption{Similar to Fig.\ref{fig3}, but showing the correlations for Solution IV and Solution V. \label{fig6}}
\end{center}
\end{figure}

In this section, we investigate the GCE solutions for the case that DM is singlino-dominated, and meanwhile $H_1$ acts as the SM-like Higgs boson.
We carry out our study in a way similar to what we did in Section III.

In Fig.\ref{fig5} we project the solutions on $\langle \sigma v \rangle_0-m_{\tilde{\chi}_1^0}$  plane (upper panel) and $\chi^2_{GCE}-m_{\tilde{\chi}_1^0}$ plane
(lower panel). For solutions marked by green lozenge, DM annihilates with the largest branching ratio into $H_1 A_1$, while for those marked by red pentastar,
DM annihilates mainly into $H_2 A_1$. In the following, we call these two kinds of solutions Solution IV and Solution V respectively. Fig.\ref{fig5} then
indicates that, for $70 {\rm GeV} \leq m_{\tilde{\chi}_1^0} \leq 87 {\rm GeV}$, Solution IV can explain the GCE quite well with the best explanation
coming from $m_{\tilde{\chi}_1^0} \simeq 81 {\rm GeV}$ with $\chi^2_{GCE} \simeq  21.4$ (corresponding to a p-value of 0.55), and for
$80 {\rm GeV} \leq m_{\tilde{\chi}_1^0} \leq 130 {\rm GeV}$, Solution V is good in accounting for the GCE with the best explanation
locating at $m_{\tilde{\chi}_1^0} \simeq 85 {\rm GeV}$ with $\chi^2_{GCE} \simeq  21.6$ and a p-value of 0.54.

Compared with the case that $H_2$ acts as the SM-like Higgs boson, we find that it is more difficult to get the GCE solutions if $H_1$ corresponds to
the SM-like Higgs boson. One important reason is that the spectrum of the singlet-dominated particles for Solution IV and V has non-trivial
requirements on the NMSSM parameters, which can not be easily satisfied due to the structure of the NMSSM itself. A good example about this argument is that
we do not find any solutions where DM mainly annihilates into $b\bar{b}$. This is due to the fact that, given a singlino-dominated DM with $30 {\rm GeV}
\leq m_{\tilde{\chi}_1^0} \leq 70 {\rm GeV}$ and meanwhile a singlet-dominated $A_1$ satisfying $m_{A_1} \simeq 2  m_{\tilde{\chi}_1^0} $, the
singlet-dominated CP-even Higgs boson is usually lighter than the SM-like Higgs boson  \cite{Cao:2014efa}.

In Table \ref{Table2}, we present detailed information for benchmark points P4, P5, P6 and P7 with points P4 and P5 belonging to Solution IV and points P6 and P7 belonging to Solution V.
This table shows that for $80 {\rm GeV} \lesssim m_{\tilde{\chi}_1^0} \lesssim 86 {\rm GeV}$, DM may annihilate into $H_1 A_1$ and $H_2 A_1$ states with
comparable rates to explain the GCE (see points P5 and P6), while for $m_{\tilde{\chi}_1^0} \simeq 78 {\rm GeV}$ ($m_{\tilde{\chi}_1^0} \simeq 120 {\rm GeV}$),
the sole annihilation channel $\tilde{\chi}_1^0 \tilde{\chi}_1^0 \to H_1 A_1$ ($\tilde{\chi}_1^0 \tilde{\chi}_1^0 \to H_2 A_1$)
can be responsible for the GCE, see point P4 (P7).

In Fig.\ref{fig6}, we show the correlations among the parameters $\lambda$, $\kappa$, $\mu$, $m_{\tilde{\chi}_1^0}$ and
$m_{A_1}$. This figure is supplement to Table \ref{Table1}, and as we will show below, it is helpful for our understanding on Solution IV and V.

\subsection{Solution IV - the $H_1 A_1$ annihilation channel}

Solution IV has the following features:
\begin{itemize}
\item The $H_1 A_1$ state must be produced close to threshold to explain the GCE, which is reflected by $\delta < 0.1 $ from our results.

\item The favored spectrum for the singlet-dominated particles is $71 {\rm GeV} \lesssim  m_{\tilde{\chi}_1^0} \lesssim 87 {\rm GeV}$,
$10 {\rm GeV} \lesssim  m_{A_1} \lesssim 40 {\rm GeV}$ and $126 {\rm GeV} \lesssim  m_{H_2} \lesssim 142 {\rm GeV}$.
Given $\kappa \sim 0.12 $ which is required to predict the right size of the annihilation $\tilde{\chi}_1^0 \tilde{\chi}_1^0
\to H_1 A_1$ for the GCE (see below), this spectrum limits parameters such as $\lambda$, $\kappa$, $\mu$ and $A_\kappa$
within rather narrow ranges, which are given in Table \ref{Table1} and also shown in Fig.\ref{fig6}.

Compared with Solution II, we find in Solution IV that, in order to predict a heavier singlet-dominated CP-even Higgs boson, the parameter
$\mu$ usually takes a larger value, $220 {\rm GeV} \lesssim \mu \lesssim 270 {\rm GeV}$. As a result, $\lambda $ must exceed about 0.6, which
can be inferred from the relation $m_{\tilde{\chi}_1^0} \simeq 2 \kappa \mu/\lambda \simeq 2 \times 0.12 \times \mu/\lambda
\simeq 80 {\rm GeV}$. This relation also suggests that $v_s \equiv \mu/\lambda \gtrsim 300 {\rm GeV} $ or $\lambda v/\mu < 0.6$, which
makes the expansions listed in Appendix feasible.

\item Similar to Solution II, the s-channel contributions to the annihilation rate $ \langle \sigma_{H_1 A_1} v \rangle_0 $ in Eq.(\ref{HIA1})
are usually much smaller than those from the $t/u$ channel, and among the $t/u$ channel contributions, the one induced by the
exchange of $\tilde{\chi}_1^0$ is dominant. However, since $H_1$ now is the SM-like Higgs boson (instead of a singlet-dominated
particle in Solution II), there still exists a slight difference between the two solutions. Explicitly speaking,
we find that each higgsino contribution to the annihilation is comparable in magnitude with the $\tilde{\chi}_1^0$  contribution,
but since the two higgsino contributions cancel each other, the total higgsino contribution is small.
These features can be explained by the following formula (see Eq.(\ref{couplings1}))
\begin{eqnarray}
&& C_{A_1 \tilde{\chi}_1^0 \tilde{\chi}_1^0} C_{H_1 \tilde{\chi}_1^0 \tilde{\chi}_1^0} \simeq -4 i \kappa^2 \frac{\lambda v}{\mu} \sin^2 \beta
\simeq -8 i \kappa^3 \frac{v}{m_{\tilde{\chi}_1^0}} \sin^2 \beta, \label{coupling-annihilation} \\
&& C_{A_1 \tilde{\chi}_1^0 \tilde{\chi}_i^0} C_{H_1 \tilde{\chi}_1^0 \tilde{\chi}_i^0} \simeq
\left \{ \begin{array}{c} \frac{i}{4} \lambda^2 \frac{\lambda v}{\mu} \sin^2 \beta, \quad for\  Higgsino-like\ \tilde{\chi}_i^0\ and  \ m_{\tilde{\chi}_i^0} < 0, \nonumber  \\
\ - \frac{i}{4} \lambda^2 \frac{\lambda v}{\mu} \sin^2 \beta, \quad for\  Higgsino-like\ \tilde{\chi}_i^0\ and  \ m_{\tilde{\chi}_i^0} > 0,  \nonumber
\end{array} \right.
\end{eqnarray}
and also by comparing Eq.(\ref{coupling-annihilation}) with equation (3.20) in  \cite{Gherghetta:2015ysa} to conclude
 that $\kappa \sim 0.12 $ is enough to predict the $\tilde{\chi}_1^0$ contributed $ \langle \sigma_{H_1 A_1} v \rangle_0 $ at the order of $10^{-26} cm^3/s$.

\item Since $m_{A_1} \lesssim 40 {\rm GeV}$ for all cases in Solution IV (see upper right panel of Fig.\ref{fig6}), the SM-like Higgs boson $H_1$
will decay into $A_1 A_1$. We checked that $Br( H_1 \to A_1 A_1 ) \lesssim 24 \%$ as required by the Higgs data.

\item We also checked that $A_1 \to b \bar{b}$ is the dominant decay mode of $A_1$, and $Br(A_1 \to \gamma \gamma ) < 5 \times 10^{-5}$ for all samples.
\end{itemize}

\subsection{Solution V - the $H_2 A_1 $ annihilation channel}

Since $H_2$ in Solution V is singlet dominated, the features of Solution V should be similar to those of Solution II. The differences mainly come from the following aspects:
\begin{itemize}
\item The spectrum of the singlet dominated particles. In Solution V, the favored spectrum is
$80 {\rm GeV} \lesssim  m_{\tilde{\chi}_1^0} \lesssim 130 {\rm GeV}$,
$18 {\rm GeV} \lesssim  m_{A_1} \lesssim 100 {\rm GeV}$ and $125 {\rm GeV} \lesssim  m_{H_2} \lesssim 146 {\rm GeV}$
with $m_{A_1} < m_{\tilde{\chi}_1^0} < m_{H_2}$ and $\delta < 0.1$. Corresponding to such a spectrum, the parameter space of Solution V
differs greatly from that of Solution II, which can be seen from Table \ref{Table1} and also from Fig.\ref{fig6}.
\item The phenomenology of some relevant particles. For example, in both Solution IV and Solution V, the favored value of $\mu$ is uplifted in comparison with
that in Solution II. As a result, the higgsino-dominated neutralinos may decay into $Z \tilde{\chi}_1^0$, which makes them
to be potentially detected at 14-TeV LHC by trilepton $+ E_T^{miss}$ signals  \cite{Cao:2014efa}.
\end{itemize}

\section{Explore the GCE solutions in future DM experiments}

\begin{figure}[t]
\begin{center}
\includegraphics[width=14cm]{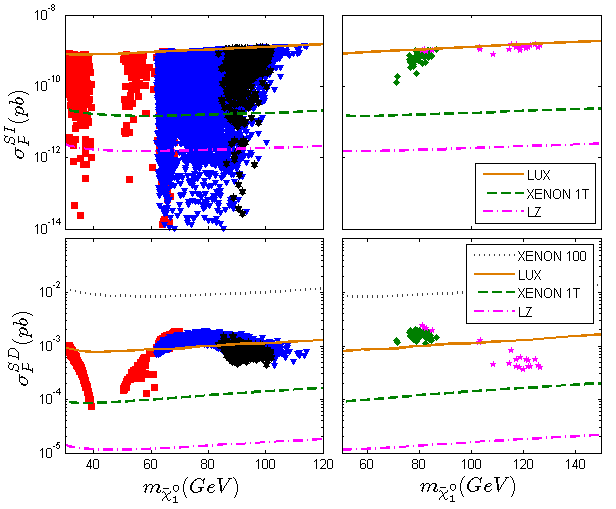}
\caption{Spin-independent (SI) and Spin-dependent (SD) cross sections for DM-nucleon scattering as a function of DM mass. Solutions in this figure are taken from Fig.\ref{fig2} and
Fig.\ref{fig5} with the same symbolic conventions. }
\label{fig7}
\end{center}
\end{figure}

In this section we investigate to what extent the GCE solutions will be explored in future DM direct detection experiments such as XENON-1T and LUX experiments \cite{future_DM_Direct_Detection}, which will improve current experimental sensitivities to DM-nucleon scattering cross sections by up to three orders. In Fig.\ref{fig7}, we project our solutions on $m_{\tilde{\chi}_1^0}-\sigma^{SI}_p$ and $m_{\tilde{\chi}_1^0}-\sigma^{SD}_p$ planes with
$\sigma^{SI}_p$ and $\sigma^{SD}_p$ denoting the spin-independent (SI) and spin-dependent (SD) cross sections respectively. The left panels in the figure are the results
for the case that $H_2$ acts as the SM-like Higgs boson, and the right panels are those for the case that $H_1$ corresponds to the SM-like Higgs boson.
The dotted lines, solid lines, dashed lines and dash dotted lines are the sensitivities
to the cross sections set by the XENON-100, LUX, XENON-1T and LZ experiments respectively. Note that so far the XENON-100 experiment has imposed constraints
on both SI and SD cross sections, while the LUX experiment only obtained limits on the SI cross section.

For $\sigma^{SI}_p$ in the $H_2$ case, we can see from Fig.\ref{fig7} that the future XENON-1T experiment is able to probe a large portion of the GCE solutions, and the
LZ experiment can test even more solutions. Anyhow, there still exist some solutions remaining untouched by these future experiments. This conclusion can be understood as follows.
In the NMSSM after considering the current experimental constraints on sfermion masses, the main contribution to $\sigma^{SI}_p$ comes from the t-channel process mediated
by the CP-even Higgses $H_{1,2}$. In this case, the Wilson coefficient $f_{q_i}$ for the operator $\bar{\tilde{\chi}}_1^0 \tilde{\chi}_1^0 \bar{q}_i q_i$ is given by \cite{Jungman}
\begin{eqnarray}
f_{q_i} \simeq  \frac{C_{H_1 \tilde{\chi}_1^0 \tilde{\chi}_1^0} C_{H_1 q_i q_i}}{2 m_{H_1}^2} +   \frac{C_{H_2 \tilde{\chi}_1^0 \tilde{\chi}_1^0} C_{H_2 q_i q_i}}{2 m_{H_2}^2}, \label{SI-cross}
\end{eqnarray}
where $C_{H_1 \tilde{\chi}_1^0 \tilde{\chi}_1^0} \simeq - \sqrt{2} \kappa (1 + 2 \lambda v/\mu)$ and $C_{H_2 \tilde{\chi}_1^0 \tilde{\chi}_1^0} \simeq 2
\sqrt{2} \kappa \lambda v/\mu$, which are given by Eq.(\ref{couplings1}) and Eq.(\ref{couplings2}) respectively.  Then Eq.(\ref{SI-cross}) indicates that, if $\kappa$ is small or
if there exists a strong cancelation between the two terms, $f_{q_i}$ or equivalently the SI cross section will be suppressed. We numerically checked
that the untouched solutions has either of the two characteristics.

On the other hand, the story for $\sigma^{SD}_p$ in the $H_2$ case is quite different. From the lower left panel of Fig.\ref{fig7} we can see that the future XENON-1T experiment
can test almost all of the GCE solutions, let alone the more sensitive LZ experiment. The underlying reason is that in the NMSSM with heavy sfermions,
the SD cross section gets contribution mainly from the $t$-channel $Z$-mediated diagram.  As a result, the size of the cross section is determined by the
$Z \tilde{\chi}_1^0 \tilde{\chi}_1^0$ coupling, which is given by
\begin{eqnarray}
g_{Z \tilde{\chi}_1^0 \tilde{\chi}_1^0} = \frac{m_Z}{\sqrt{2} v} ( N_{13}^2 - N_{14}^2 ) \simeq - \frac{m_Z}{\sqrt{2} v} \frac{\lambda^2 v^2}{\mu^2} ( 1 - \frac{4 \kappa^2}{\lambda^2} )
\simeq - \frac{m_Z}{\sqrt{2} v} \frac{\lambda^2 v^2}{\mu^2} ( 1 - \frac{m_{\tilde{\chi}_1^0}^2}{\mu^2} ).
\end{eqnarray}
In getting this expression, we have used the approximations for $N_{13}$, $N_{14}$ and $m_{\tilde{\chi}_1^0}$.
Then from the results presented in Fig.\ref{fig3}, one can infer that except for some rare cases
of Solution I, the SD cross section is not suppressed too much.

In a similar way, one can analyze the results for the $H_1$ case, which are shown on the right panels of Fig.\ref{fig7}. For example, the upper right panel indicates that the SI cross
sections in Solution IV and Solution V are usually larger than $10^{-10} pb$. This may be understood by a weak cancelation between the two terms in Eq.(\ref{SI-cross}).
Compared with the $H_2$ case, both the SI cross section and the SD cross section in the $H_1$ case are large and consequently, all the solutions will be tested
by XENON-1T experiment.

In principle, the GCE solutions in the NMSSM may also be tested by electroweakino production processes at the LHC  \cite{Cao:2014efa}.
We will discuss such an issue in our forthcoming work.

\section{Summary}

In this work, we took into account the recently reported foreground and background uncertainties for the GCE and
investigated its explanation by DM annihilation in the framework of the NMSSM. We carried out a sophisticated scan over
the NMSSM parameter space by considering various experimental constraints such as the Higgs data, $B-$physics observables,
DM relic density, LUX experiment and the dSphs constraints. Then for each surviving parameter point we performed a fit to the GCE
spectrum by using the correlation matrix that incorporated both the statistical and systematic uncertainties of the measured
excess. Our results indicate that due to the introduction of the gauge singlet Higgs superfield, the NMSSM with a singlino-dominated
DM has multiple DM annihilation channels that are able to explain the GCE quite well, and all of these explanations require
the singlet-dominated particles (including one neutralino, one CP-even and one CP-odd Higgs bosons) to be moderately light.
We also discussed to what extent the future DM direct detection experiments can explore the GCE solutions, and
we conclude that the XENON-1T experiment is very promising in
testing nearly all the solutions.

When choosing the scenario of particle spectrum, we focused on a singlino-dominated DM and considered the cases that either $H_2$ or $H_1$ acts as the SM-like Higgs boson.
For the popular situation that $H_2$ corresponds to the SM-like Higgs, we have the following observations on the GCE solutions:
\begin{itemize}
\item  The pure DM annihilation channel $\tilde{\chi}_1^0 \tilde{\chi}_1^0 \to b \bar{b} $ or $\tilde{\chi}_1^0 \tilde{\chi}_1^0 \to H_1 A_1 $
can provide a good fit to the GCE spectrum, while the channel $\tilde{\chi}_1^0 \tilde{\chi}_1^0 \to W^+ W^- $ must mix sizeably with the channel
$\tilde{\chi}_1^0 \tilde{\chi}_1^0 \to H_1 A_1 $ to account for the GCE.

\item For the annihilation $\tilde{\chi}_1^0 \tilde{\chi}_1^0 \to b \bar{b}$,  DM mass is now allowed in the range from $30 {\rm GeV}$ to
$70 {\rm GeV}$ which is much wider than before. With the help of an appropriate s-channel resonance, the singlet trilinear self-coupling
parameter $\kappa$ can be as low as 0.02 to explain the GCE. Moreover, the higgsino mass parameter $\mu$ is upper bounded by about $300 {\rm GeV}$
to ensure a correct DM relic density. Since there exist strong correlations between independent parameters, such an explanation suffers from a fine
tuning problem, which is usually less than $1\%$.

\item The annihilation $\tilde{\chi}_1^0 \tilde{\chi}_1^0 \to H_1 A_1$ may provide a better explanation than the channel
$\tilde{\chi}_1^0 \tilde{\chi}_1^0 \to b \bar{b}$ when $H_1A_1$ is produced close to threshold, and the best interpretation corresponds to
a p-value of 0.55. In this kind of explanation, the singlet-dominated particles must satisfy
$60 {\rm GeV} \lesssim  m_{\tilde{\chi}_1^0} \lesssim 115 {\rm GeV}$, $10 {\rm GeV} \lesssim  m_{A_1} \lesssim 110 {\rm GeV}$,
$60 {\rm GeV} \lesssim  m_{H_1} \lesssim 120 {\rm GeV}$ and $\delta < 0.2$. This imposes non-trivial
constraints on the NMSSM parameters, especially that $\mu$ must be less than about $200 ~{\rm GeV}$. Among various contributions to the annihilation,
the dominant one comes from the $\tilde{\chi}_1^0$-contributed $t/u$ channel diagrams, in which the parameter $\kappa$ plays an important role
in deciding the annihilation rate.

\item Apart from the necessary mixing with the $H_1 A_1 $ final states, $W^+W^-$ pair in the annihilation $\tilde{\chi}_1^0 \tilde{\chi}_1^0 \to W^+ W^- $
must be produced close to threshold to account for the GCE. A small $\mu$ less than about 150 GeV is necessary to increase the annihilation rate
through the $t/u$-channel contributions induced by a higgsino-dominated chargino. The LHC search for trilepton $+ E_{T}^{miss}$ signal
can not exclude such a possibility since the electroweakino production rates at the LHC are relatively low, and meanwhile since the splitting between
$\mu$ and $m_{\tilde{\chi}_1^0}$ is compressed.

\item The detection of spin-independent scattering in the future XENON-1T and LUX experiments are able to cover a large portion of the GCE-favored
parameter space, while the spin-dependent detection have a stronger potential to test nearly all of the relevant parameter region.
\end{itemize}

As for the case that $H_1$ acts as the SM-like Higgs boson, the features of the GCE solutions are quite different, which are as follows:
\begin{itemize}
\item  In comparison with the $H_2$ case, it is difficult to find GCE solutions when $H_1$ corresponds to the SM-like Higgs boson, and especially we
did not find any solution that DM annihilates mainly into $b\bar{b}$. The reason is, assuming $H_1$ to be the SM-like Higgs boson,
there must exist sizeable mass splittings among the light singlet-dominated particles to explain the GCE, which is difficult to realize in the NMSSM
due to the theoretical structure itself.

\item For $80 {\rm GeV} \lesssim m_{\tilde{\chi}_1^0} \lesssim 86 {\rm GeV}$, DM may annihilate into $H_1 A_1$ and $H_2 A_1$ states with
comparable rates to explain the GCE, while for $m_{\tilde{\chi}_1^0} \lesssim 80 {\rm GeV}$ ($m_{\tilde{\chi}_1^0} \gtrsim 100 {\rm GeV}$),
the sole annihilation $\tilde{\chi}_1^0 \tilde{\chi}_1^0 \to H_1 A_1$ ($\tilde{\chi}_1^0 \tilde{\chi}_1^0 \to H_2 A_1$) can be responsible
for the GCE. For all these solutions, the singlet-dominated particle $H_2$ and the parameter $\mu$ must satisfy
$125 {\rm GeV} \lesssim m_{H_2} \lesssim 145 {\rm GeV}$ and $210 {\rm GeV} \lesssim \mu \lesssim 270 {\rm GeV}$.

\item Both the spin-independent and spin-dependent detection in the future XENON-1T experiment have a great potential to test the relevant parameter space.
\end{itemize}

Before we end our discussion, we would like to comment briefly on the interpretation of the GCE with a bino-like DM. Like the singlino-dominated DM case,
a light $A_1$ with mass below about $140 {\rm GeV}$ is necessary for such a work, and this $A_1$ prefers to be singlet-dominated \footnote{In the NMSSM,
a light $A_1$ with mass below about $100 {\rm GeV}$ may have a large doublet component
if the elements of the CP-odd Higgs mass matrix satisfy
$\mathcal{M}^2_{P,22} \gg \mathcal{M}^2_{P,12} \gg \mathcal{M}^2_{P,11}$ (see benchmark points P3 and P4 in \cite{Cao:2014kya})
or $\mathcal{M}^2_{P,22} \simeq \mathcal{M}^2_{P,11} \sim \mathcal{M}^2_{P,12}$ (see the point presented in
Table \ref{Table2} of \cite{Kozaczuk:2015bea}). In either case, $m_{A_1}$ should be significantly smaller
than $m_{H^\pm}$ to escape experimental constraints. Previous studies have suggested that a light
doublet-dominated $A_1$ might also explain the galactic center excess. However, due to the requirements on the elements
this scenario occurs only in specific portions of the parameter space and is significantly more experimentally
constrained than those we considered. In fact, in our scans for the GCE we did not find any parameter points
with the doublet component of the light $A_1$ exceeding 0.1. In summary, a light doublet-dominated $A_1$ may exist,
as suggested by e.g. Ref.~\cite{Kozaczuk:2015bea}, but it is  fair to say that without a very delicate
parameter tuning, it is difficult to obtain in explaining the GCE, especially when one considers
more constraints than previous literatures. About this conclusion, we thank the authors of
\cite{Kozaczuk:2015bea} for helpful discussion.}. The difference is that, for the bino-like DM case,
the interaction of the DM with $A_1$ is relatively small and consequently the annihilation
$\tilde{\chi}_1^0 \tilde{\chi}_1^0 \to H_i A_1$ can not explain the GCE any more due to its rather
low annihilation rate. Also due to the suppressed
interaction, $m_{A_1}$ must be closer to $2 m_{\tilde{\chi}_1^0}$ for the annihilation $\tilde{\chi}_1^0 \tilde{\chi}_1^0 \to b \bar{b}$
to account for the GCE, and thus the theory has to be tuned in a more elaborated way. Our sophisticated scan over the relevant NMSSM parameter space
verified these conclusions.

\section*{Acknowledgement}

This work was supported by the National Natural Science Foundation
of China (NNSFC) under grant No. 10821504, 11222548, 11121064, 11135003, 90103013
and 11275245, and by the CAS Center for Excellence in Particle Physics (CCEPP).

\appendix

\section{Properties of the singlet-dominated particles}

In this appendix, we present some analytic expressions for the masses and couplings of the singlet dominated particles,
such as $\tilde{\chi}_1^0$ and $A_1$ in the NMSSM. These expressions are obtained by diagonalizing the mass matrices
of the particles (like done in  \cite{Cheung:2014lqa}), and are good approximations in certain cases. They are helpful in understanding
the results presented in this work. In the following, we will follow notations and conventions
consistent with \cite{NMSSM_review_0910.1785} for the $Z_3$ NMSSM.

\subsection{Neutralino masses and mixings}

In the basis $\psi^0 = (-i \lambda_1, - i \lambda_2^3, \psi_d^0, \psi_u^0, \psi_S)$, the neutralino mass matrix is:
\begin{equation}
{\cal M} = \left(
\begin{array}{ccccc}
M_1 & 0 & -\frac{g_1 v_d}{\sqrt{2}} & \frac{g_1 v_u}{\sqrt{2}} & 0 \\
  & M_2 & \frac{g_2 v_d}{\sqrt{2}} & - \frac{g_2 v_u}{\sqrt{2}} &0 \\
& & 0 & -\mu & -\lambda v_u \\
& & & 0 & -\lambda v_d\\
& & & & \frac{2 \kappa}{\lambda} \mu
\end{array}
\right). \label{eq:MN}
\end{equation}
If the bino and wino fields are decoupled, the mass eigenstates of the neutralinos
can be approximated by
\begin{eqnarray}
\tilde{\chi}_1^0 & \approx & N_{13} \psi_d^0 + N_{14}\psi_u^0 + N_{15} \psi_S, \nonumber \\
\tilde{\chi}_i^0 & \approx & N_{i3} \psi_d^0 + N_{i4}\psi_u^0 + N_{i5} \psi_S,
\end{eqnarray}
where $\tilde{\chi}_1^0$ denotes the lightest neutralino with $\psi_S$ field as its dominant component in this work,
and $\tilde{\chi}_i^0$ represents a higgsino-like neutralino.

In the limit of $|\mu| \gg \lambda v$, $1 \gg \kappa/\lambda $ and $\tan \beta \gg 1$, one can expand the neutralino
masses and $N_{ij}$ by the power of  $\lambda v/\mu \equiv v/v_s$ to get the following approximations:
\begin{eqnarray}
m_{\tilde{\chi}_1^0} & \approx & \frac{2 \kappa}{\lambda} \mu + \frac{\lambda^2 v^2}{\mu^2} ( \mu \sin 2 \beta - \frac{2 \kappa}{\lambda} \mu ), \nonumber \\
\frac{N_{13}}{N_{15}}&=& \frac{\lambda  v }{\mu ^2-m_{\tilde{\chi}_1^0}^2} \cos \beta \left(\tan \beta m_{\tilde{\chi}_1^0}-\mu \right) \approx \frac{2 \kappa  v }{\mu} \sin \beta, \nonumber \\
\frac{N_{14}}{N_{15}}&=&\frac{-\lambda  v }{\mu ^2-m_{\tilde{\chi}_1^0}^2} \sin \beta \left(\mu -\frac{m_{\tilde{\chi}_1^0}}{\tan \beta}\right) \approx - \frac{\lambda  v }{\mu} \sin \beta,  \nonumber \\
N_{15} &=&\left(1+ \frac{N^2_{13}}{N^2_{15}}+\frac{N^2_{14}}{N^2_{15}}\right)^{-1/2} \approx 1, \quad N_{i3}  \approx  \frac{1}{\sqrt{2}} Sgn(m_{\tilde{\chi}_j^0}) \theta(m_{\tilde{\chi}_j^0}), \nonumber \\
N_{i4} & \approx & - \frac{1}{\sqrt{2}} Sgn(\mu) \theta(m_{\tilde{\chi}_j^0}), \quad
N_{i5}  \approx  - \frac{1}{\sqrt{2}} \frac{\lambda v \sin \beta}{\mu} Sgn(\mu) \theta(m_{\tilde{\chi}_j^0}). \label{Basic-Approximation}
\end{eqnarray}
In above expressions, the $Sgn$ and $\theta$ functions are defined by
\begin{eqnarray}
Sgn(x) =  \left \{ \begin{array}{c} 1 \quad \ \  if \ x \ge 0, \\
-1 \quad if \ x < 0,
\end{array} \right.  \quad \theta(x) =  \left \{ \begin{array}{c} 1 \quad \ if \ x \ge 0, \\
i \quad \ if \ x < 0.
\end{array} \right.
\end{eqnarray}

Likewise, one may consider the case that the wino and the singlino fields decouple. In this case, the mass
eigenstates of the neutralinos is approximated by
\begin{eqnarray}
\tilde{\chi}_1^0 & \approx & N_{11} (-i \lambda_1) + N_{13} \psi_d^0 + N_{14}\psi_u^0, \nonumber \\
\tilde{\chi}_i^0 & \approx & N_{i1} (-i \lambda_1) + N_{i3} \psi_d^0 + N_{i4}\psi_u^0.
\end{eqnarray}

In the limit of $\tan \beta \gg 1$, $|\mu| \gg g_2 v_u$ and $|\mu| \gg M_1$, we have the following approximations:
\begin{eqnarray}
m_{\tilde{\chi}_1^0} & \approx & M_1 - \frac{m_Z^2 \sin^2 \theta_W}{\mu^2} (\mu \sin 2 \beta + M_1 ), \nonumber \\
\frac{N_{13}}{N_{11}} &\approx & \frac{m_Z \sin \theta_W }{\mu} \sin \beta, \nonumber \\
\frac{N_{14}}{N_{11}} &\approx & - \frac{m_Z \sin \theta_W }{\mu} \cos \beta (1 + \tan \beta \frac{M_1}{\mu}),  \nonumber \\
N_{11} &=&\left(1+ \frac{N^2_{13}}{N^2_{11}}+\frac{N^2_{14}}{N^2_{11}}\right)^{-1/2} \approx 1, \quad N_{i3}  \approx  \frac{1}{\sqrt{2}} Sgn(m_{\tilde{\chi}_j^0}) \theta(m_{\tilde{\chi}_j^0}), \nonumber \\
N_{i4} & \approx & - \frac{1}{\sqrt{2}} Sgn(\mu) \theta(m_{\tilde{\chi}_j^0}), \quad
N_{i1}  \approx   \frac{1}{\sqrt{2}} \frac{m_Z \sin \theta_W \sin \beta}{\mu} Sgn(\mu) \theta(m_{\tilde{\chi}_j^0}).
\end{eqnarray}

\subsection{CP-odd Higgs mass matrix}

In the $(A,S_I)$ ``interaction'' basis, the mass matrix for CP-odd Higgs bosons is given by
\begin{eqnarray}
\mathcal{M}^2_P & = & \left(
\begin{array}{cc}
m_A^2
& \lambda v ( \frac{m_A^2}{2 \mu} \sin 2 \beta - \frac{3 \kappa \mu}{\lambda} )
\\
& \lambda^2 v^2 \sin 2 \beta ( \frac{m_A^2}{4 \mu^2} \sin 2 \beta + \frac{3 \kappa}{2 \lambda}) - \frac{3 \kappa A_\kappa \mu}{\lambda} \\
\end{array}
\right).
\end{eqnarray}
In the case of $m_A \gg max(v,|A_\kappa|,|\mu|)$, $\kappa/\lambda \ll 1$ and $\tan \beta \gg 1$, the lighter CP-odd scalar
$A_1$ is singlet dominated with its squared mass given by
\begin{eqnarray}
m_{A_1}^2 \approx \frac{9}{2} \lambda \kappa v^2 \sin 2 \beta - \frac{3 \kappa A_\kappa \mu}{\lambda}.  \label{A1-mass}
\end{eqnarray}
This approximation indicates that, without considering the radiative corrections, the singlet-dominated CP-odd scalar mass
is determined by the parameters $\lambda$, $\kappa$, $\mu$ as well as $A_\kappa$. The components of  $A_1$ can be written as
\begin{eqnarray}
\frac{P_{A_1,A}}{P_{A_1,S_I}} &\approx& -\frac{\lambda  v}{m_A^2} \left( \frac{m_A^2}{2\mu} \sin 2\beta -3 \frac{\kappa  \mu }{\lambda } \right) \approx -\frac{\lambda v}{2 \mu} \sin 2 \beta, \nonumber \\
P_{A_1,S_I} &=&\left(1+\frac{P_{A_1,A}^2}{P_{A_1,S_I}^2}\right)^{-1/2} \approx 1,
\end{eqnarray}
where $P_{A_1,A}$ is the active component and $P_{A_1,S_I}$ is the singlet component of the $A_1$.

\subsection{CP-even Higgs mass matrix}
\label{sec:CPE}

In the basis $\left( S_1= \cos \beta h_u - \sin \beta h_d, S_2=  \sin \beta h_u + \cos \beta h_d, S_3 = h_S \right)$, the mass matrix elements for
the CP-even scalars are \cite{Cao-NMSSM}
\begin{eqnarray}
{\cal M}^2_{11}&=&M_A^2+(m_Z^2-\lambda^2v^2)\sin^22\beta,\\
{\cal M}^2_{12}&=&-\frac{1}{2}(m_Z^2-\lambda^2v^2)\sin4\beta,\\
{\cal M}^2_{13}&=&-(M_A^2\sin2\beta+\frac{2\kappa\mu^2}{\lambda})\frac{\lambda v}{\mu}\cos2\beta,\\
{\cal M}^2_{22}&=&m_Z^2\cos^22\beta+\lambda^2v^2\sin^22\beta,\label{22}\\
{\cal M}^2_{23}&=& 2 \lambda \mu v \left[1 - (\frac{M_A \sin 2\beta}{2 \mu} )^2
-\frac{\kappa}{2 \lambda}\sin2\beta\right],\\
{\cal M}^2_{33}&=& \frac{1}{4} \lambda^2 v^2 (\frac{M_A \sin 2\beta}{\mu})^2
+ \frac{\kappa\mu}{\lambda} (A_\kappa +  \frac{4\kappa\mu}{\lambda} )
 - \frac{1}{2} \lambda \kappa v^2 \sin 2 \beta, \label{33}
\end{eqnarray}
where $S_2$ is nothing but the Higgs field in the SM, ${\cal M}^2_{22}$ is its mass at tree level without considering the mixing
among $S_i$, and the second term $\lambda^2v^2\sin^22\beta$ in ${\cal M}^2_{22}$ originates from the coupling
$\lambda\hat{H_u} \cdot \hat{H_d} \hat{S}$ in the superpotential.

The mass eigenstates $H_i$ are defined by
\begin{eqnarray}
H_i = V_{i1} S_1 + V_{i2} S_2 + V_{i3} S_3,
\end{eqnarray}
where $V$ is the rotation matrix to diagonalize the mass matrix. For the $S_2$-dominated mass eigenstate $H_j$,
current Higgs data have required it to be highly SM-like, i.e. $V_{j,1}, V_{j,3} \ll 1$, so in the case of the hierarchy structure
${\cal M}^2_{11} \gg max({\cal M}^2_{22}, {\cal M}^2_{33})$, ${\cal M}^2_{23} \ll |{\cal M}^2_{22} - {\cal M}^2_{33}|$. If we
decouple the MSSM-like heavy Higgs, $S_1$, from the other two,
the $2 \times 2$ reduced mass matrix in the $(S_2,S_3)$ basis is given by \cite{Cao-NMSSM}:
\begin{eqnarray}
\mathcal{M}^2_{S_2S_3} & = & \left(
\begin{array}{ccc}
{\cal M}^2_{22}
& {\cal M}^2_{23}
\\ & {\cal M}^2_{33} - \frac{\lambda ^2 v^2 m_A^2}{16 \mu ^2} \sin^2 4 \beta -\frac{\kappa ^2 \mu ^2 v^2 }{m_A^2}\cos^2 2 \beta
   - \lambda \kappa  v^2 \cos^2 2 \beta \sin 2 \beta
\end{array}
 \right)
\end{eqnarray}
The (2,2) element of the reduced ($2\times 2$) matrix, which in the limit of zero-mixing with the other Higgs should give
singlet scalar mass in the $Z_3$ NMSSM,  is given by:

\begin{eqnarray}
\mathcal{M}^2_{S_2 S_3}(2,2)&=&\frac{\kappa  \mu
  }{\lambda } \left(A_{\kappa }+\frac{4 \kappa  \mu }{\lambda }\right)+\frac{\lambda ^2 v^2 m_A^2}{4 \mu ^2} \left(1- \cos^2 2 \beta  \right) \sin^2 2 \beta \nonumber \\
  && -\frac{\kappa ^2 \mu ^2 v^2 }{m_A^2} \cos^2 2 \beta -\frac{1}{2} \kappa  \lambda  v^2 \left(2 \cos^2 2 \beta + 1\right) \sin 2 \beta.
\end{eqnarray}

Setting $ {\cal M}^2_{23} \sim 0$,  i.e. $m_A^2 = \frac{4 \mu^2}{\sin^2 2\beta} ( 1 - \frac{\kappa}{2\lambda} \sin 2 \beta )$, and taking $\tan \beta \gg 1$, we have
\begin{eqnarray}
\mathcal{M}^2_{S_2 S_3 }(2,2)& \approx &\frac{\kappa  \mu
  }{\lambda } \left(A_{\kappa }+\frac{4 \kappa  \mu }{\lambda }\right)
\end{eqnarray}
This approximation indicates again that, without considering the radiative corrections, the singlet-dominated CP-even scalar mass
is determined by the parameters $\lambda$, $\kappa$, $\mu$ and $A_\kappa$.

\subsection{Some properties of the singlet-dominated particles}

With the assumptions that $M_A \gg max(|\mu|, |A_\kappa|)$, $|\mu| \gg \lambda v$, $\tan \beta \gg 1$ and $\kappa/\lambda \ll 1$, one
can approximate the masses and couplings of the singlet dominated particles, such as $\tilde{\chi}_1^0$ and $A_1$,
by simple analytic expressions \cite{Cheung:2014lqa}. In the following, we list some of the coupling expressions used in our discussion,
which are denoted by $C_{XYZ}$ hereafter. These expressions are actually expand the corresponding exact ones by the power of $\lambda v/\mu$.
\begin{eqnarray}
C_{A_1 b \bar{b}} & = & \frac{i m_b \tan \beta}{\sqrt{2} v} P_{A_1 A} \approx - \frac{i m_b}{\sqrt{2} v} \frac{\lambda v}{\mu},  \nonumber \\
C_{A_1 \tilde{\chi}_1^0 \tilde{\chi}_j^0} & \approx & \left \{ \begin{array}{l} - i \sqrt{2} \kappa ( 1 + 2 \frac{\lambda v}{\mu}  ) \quad \quad
\quad \ \ \ for \ j= 1, \\
- \frac{i \lambda}{2} \frac{\lambda v}{\mu} Sgn(m_{\tilde{\chi}_j^0}) \theta(m_{\tilde{\chi}_j^0})  \quad for \ higgsino-like\ \tilde{\chi}_j^0,
\end{array} \right.  \nonumber \\
C_{H_i \tilde{\chi}_1^0 \tilde{\chi}_j^0} & = & - i C_{A_1 \tilde{\chi}_1^0 \tilde{\chi}_j^0} \quad \quad \quad \quad \quad \quad \quad \ \ if\ H_i \ is \ singlet\ dominated.  \label{couplings1}
\end{eqnarray}

Likewise, if $H_i$ is the SM-like Higgs boson, we have
\begin{eqnarray}
C_{H_i \tilde{\chi}_1^0 \tilde{\chi}_j^0} & \approx & \left \{ \begin{array}{l} 2 \sqrt{2} \kappa \frac{\lambda v}{\mu}  \quad \quad
\quad \quad \quad \quad \ \  for \ j= 1, \\
\frac{\lambda}{2} Sgn(m_{\tilde{\chi}_j^0}) \theta(m_{\tilde{\chi}_j^0})  \quad \quad  for \ higgsino-like\ \tilde{\chi}_j^0,
\end{array} \right.  \label{couplings2}
\end{eqnarray}

\end{document}